\documentclass[mypaper,7pt,twoside]{CoAst}
\usepackage{epsf,graphicx,fancyhdr}
\usepackage{amssymb,verbatim,nicefrac} 
\renewcommand{\chaptermark}[1]{\markboth{#1}{}}

\newcommand{\teff}{\ensuremath{T_{eff}}}             
\newcommand{\logg}{\ensuremath{\log g}}                     

\newcommand{\kopf}{\small\itshape Comm. in Asteroseismology\\ Vol. 144, 2004}
\newcommand{\Authors}[1]{\begin{center}\normalsize\bf\sf #1 \end{center}}

\renewcommand{\author}[1]{\begin{center}\normalsize\bf\sf #1 \end{center}}
\newcommand{\Address}[1]{\begin{center}\small\sf #1 \end{center}}

\renewenvironment{abstract}{\section*{Abstract}\normalsize\sf}{}
\newcommand{\References}[1]{\begin{flushleft}{\large References\\}\vspace*{2mm}\small #1 \end{flushleft}}

\newcommand{\chapterDSSN}[2]{\chapter[\sf\normalsize #1\\ \footnotesize \hspace*{5mm}by #2 \sf\normalsize][]{#1\\}\rhead[\fancyplain{}{\sf\footnotesize \center{#1}}]{\fancyplain{}{\sffamily\thepage}}\lhead[\fancyplain{\kopf}{\sffamily\thepage}]{\fancyplain{\kopf}{\sf\footnotesize \center{#2}}}}

\newcommand{\figureDSSN}[5]{\begin{figure}[#4]
\centering
\includegraphics*[#5]{#1}
\caption{#2}
\label{#3}
\end{figure}}

\renewcommand{\chaptername}{}
\newcommand{\acknowledgments}[1]{\vspace*{5mm}\noindent\begin{bf}Acknowledgments. \end{bf} #1}

\newcommand{\nc}{\newcommand}
\nc{\rnc}{\renewcommand}

\nc {\atlas} {{\tt ATLAS9} }
\nc {\smgt}  {{\tt SMGT}}
\nc {\nemo}  {{\tt NEMO}}

\nc {\logtau}{\ensuremath{\log \tau_{\rm Ross}} } 
\nc {\mh}    {\ensuremath{[M/H]}}
\rnc{\teff}  {\ensuremath{T_{\rm eff}}}
\nc {\vmic}  {\ensuremath{v_{\rm micro}}} 

\nc {\cp}    {\clearpage}
\nc {\nn}    {\nonumber}
\nc {\ul}    {\underline}
\nc {\vl}    {\vline}
\nc {\vlx}[1]{\hspace{-2.2mm}\vl\hfill\hspace{2.1mm} #1 \hfill\mbox{}}

\nc {\ub} {\underbrace}
\nc {\be} {\begin{equation}}
\nc {\ee}   {\end{equation}}
\nc {\bea}{\begin{eqnarray}}
\nc {\eea}  {\end{eqnarray}}

\nc {\fDelt} [2]{\frac{\Delta_{#1}}{\Delta_{#2}}} 
\nc {\ffDelt}[4]{\fDelt{#1}{#2}\fDelt{#3}{#4}}    

\newlength{\hspl}                
\nc       {\hspx}[1]             
          {\settowidth{\hspl}{#1}%
           \hspace{\hspl}}
\nc {\hm} {\hspx{$-$}}           

\nc {\ri} {\\\hspace{7mm}}       

\begin{document}
\sf

\chapterDSSN
{Interpolation of Stellar Model Grids and Application to the \nemo\ Grid}
{J.~Nendwich, U.~Heiter, F.~Kupka, N.~Nesvacil and W.W.~Weiss}


\Authors{J.~Nendwich$^1$, U.~Heiter$^2$, F.~Kupka$^3$, N.~Nesvacil$^{4,1}$, 
         W.W.~Weiss$^1$} 
\Address{
$^1$ Institut f\"ur Astronomie, T\"urkenschanzstrasse 17, 1180 Vienna,
     Austria\\
$^2$ Department of Astronomy and Space Physics, Uppsala University, Box 515,
     SE-75120 Uppsala, Sweden\\
$^3$ Max-Planck-Institute for Astrophysics, Karl-Schwarzschild Str.\ 1,
     85741 Garching, Germany\\
$^4$ European Southern Observatory (ESO), Alonso de Cordova 3107, Vitacura,
     Santiago, Chile
}

\begin{abstract}
\nemo{\tt.2003}\footnote{Vienna \ul{Ne}w \ul{Mo}del Grid of Stellar
Atmospheres} is a DVD with 91,520 stellar model atmospheres representing a 5D
grid of modified \atlas atmospheres; the purpose of the modifications was to
include different treatments of convection and higher vertical resolution. In
addition, for every model fluxes are provided and color indices for 14
different photometric systems. Because the model grid contained gaps due to
non-converging models, we developed and applied a set of  4D interpolation
routines to complete the grid.\\ All the data, which will continuously be
updated, can be found on the \nemo\ homepage {\tt
http://ams.astro.univie.ac.at/nemo/} and are available via DVD.
\end{abstract}


\section*{Introduction}

As described in Heiter et al.~(2002, hereafter {\tt H02}) several new sets of
grids of model stellar atmospheres were computed with modified versions of the
\atlas code. The individual sets differed from each other, and from previous
ones, essentially in the physics used for the treatment of the convective
energy transport, in the higher vertical resolution of the atmospheres, and in
a finer grid in the (\teff, \logg) plane. The improvements related to
resolution enabled the computation of derivatives of color indices, as well as
limb darkening coefficients and derivatives (Garrido et al.\ 2001, 2002; Barban
et al.\ 2003), accurate enough for pulsation mode identification, and of smooth
pre-main sequence evolutionary tracks (Montalb\'an et al.\ 2004). A description
of the modifications of \atlas with respect to their treatment(s) of convection
was also provided by {\tt H02}. While the grids published by Kurucz (1993b
({\tt K93b}), 1998) and Castelli et al.~(1997) favor a rather efficient
convection in stellar atmospheres due to their calibration based on solar
central intensities, the new sets of grids were computed for convection models
which predict a much lower efficiency, at least within the photosphere. The
motivation behind this has been to allow a choice between the two assumptions,
as none of the presently implemented local, homogeneous models are able to
reproduce all observations, whether for the sun or for groups of stars spread
over wider regions of the HR diagram (see, e.g., Smalley et al.\ 2002). In
addition, model atmospheres with lower convective efficiency than those in
Kurucz ({\tt K93b}, 1998) and Castelli et al.~(1997) distributions allow
recovering a larger number of observations for a given choice of convection
model and its parameters ({\tt H02}, D'Antona et al.\ 2002, D'Antona \&
Montalb\'an 2003, Montalb\'an et al.\ 2004, Stassun et al.\ 2004). Schuler et
al.\ (2004) used different sets of \atlas grids, including the 72 layer models
presented here, to derive oxygen abundances in open cluster dwarfs. Their work
shows that the differences in abundances due to different models are small
compared to the typical error in the relative oxygen triplet abundances.


\section*{Parameter range}

As discussed in {\tt H02}, the computations for all values of the grid
parameters have been finished and we can present the whole 5D grid (Table
\ref{tabRange}).

\begin{table}[t]
 \begin{tabular}{|l|c|@{}c@{}|*{3}{c|}}
  \hline
  Parameter & From   & To   & \begin{tabular}{c}Increment\\
                              (List of Values)\end{tabular}
                                            & Unit    & \#\\
  \hline\hline
  \teff     & 4,000  & $\begin{array}{c}10,000\\
                           \hspace{3.5mm}8,600^a\end{array}$
                            & 200           & K       & $\begin{array}{c}31\\
                                                \hspace{1.7mm}24^a\end{array}$\\
  \hline
  \logg     & 2.0    & 5.0  & 0.2           & \{cgs\}$^b$ & 16 \\
  \hline
            &        &      & ($-$2.0, $-$1.5, $\pm$1.0, & & \\
  \mh       & $-$2.0 & +1.0 & \ $\pm$0.5, $\pm$0.3, $\pm$0.2,
                                            & dex     & 13 \\
            &        &      & $\pm$0.1, 0.0)
                                            &         &     \\
  \hline
  \vmic     & 0      & 4    & (0, 1, 2, 4)  & km$\,$s$^{-1}$    & 4 \\
  \hline
  Convec.   & \multicolumn{4}{c}
                {\hfill CGM($\alpha^*$=0.09) / 72, \hfill
                        MLT($\alpha$=0.5)    / 72, \hfill\ }\vline &  \\
  /resol.   & \multicolumn{4}{c}
                {\hspace{1.85mm} CGM($\alpha^*$=0.09) / 288, \hspace{0.2mm}
                                 CM                   / 288 \hfill\ }\vline
                                                  & \raisebox{2mm}[0mm][0mm]4 \\
  \hline
  \multicolumn{5}{c}{\vlx{Total number of model atmospheres}}\vline 
                                                     & 91,520 \\
  \hline
 \end{tabular}
 \caption{Parameter range of the  grid\newline
          `Convec.' = convection model, `resol.' = vertical resolution (number
          of layers)\newline
          $^a$ for the non CGM models, see also text\newline
          $^b$ $\log$ cm s$^{-2}$}
 \label{tabRange}
\end{table}

The convection models CM (Canuto \& Mazzitelli, 1991) and CGM (Canuto, Goldman
\& Mazzitelli, 1996) and their implementation into \atlas are summarized in {\tt
H02}. As is shown there as well, the convective flux can be neglected for 
temperatures higher than 8\,600 K. Thus, convection has been turned off for such
models and they are computed only once for models with 288 layers and once for
models with 72 layers and are included in the CGM subgrids. This explains the
two different temperature ranges in Table \ref{tabRange} and the total number
of models of 91,520. We also repeat here that the uppermost layers of our
models are located at \logtau = $-6.875$ and the difference of consecutive
layers in \logtau is $0.125$ and $0.03125$ for models with 72 and 288 layers,
respectively.\footnote{The lowermost layer is located at \logtau $= -6.875 +
(72-1)/8 = +2.0$ and $-6.875 + (288-1)/32 = +2.09375$ for models with 72 and
288 layers, respectively.} For all models, the opacity distribution functions
calculated by Kurucz (1993a) were used.


\section*{Colors}
\label{secColors}

Colors and color indices in 14 photometric systems have been calculated with
the program "colors", which is based on the programs of Kurucz CD-ROM 13 ({\tt
http://kurucz.harvard.edu/programs/COLORS/}) revised and rewritten in Fortran90
by JN. This program is also available on the \nemo\ website as well as on DVD.
The color indices for specific photometric systems have been normalized using
model fluxes and measured indices for standard stars as follows: $\beta$~Leo
(\teff=8\,850, \logg=4.16, \mh=0.0, \vmic=2) for Walraven VBLUW, HD 83373
(9\,250, 4.00, +0.0, 2) for the {\itshape hk194} index of Str\"omgren uvbyCa, a
B star (30\,000, 4.00, +0.0, 2) for Vilnius, and Vega (9\,550, 3.95, $-$0.5, 2)
for Cousins (Cape) VRI, $\Delta$a, Geneva, JHKcit, (V)RIJKL, thirteen, and UBV.
DDO and HST color indices have not been normalized. The standard star colors
and resulting zero points, as well as the references for the filter
transmission curves that have been used, are given in Tables \ref{tabBIG1} -
\ref{tabBIG3}. Colors in the DDO system were calculated with data from Lamla
(1982, {\tt L82}). The HST filter curves were taken from {\tt
ftp://ftp.stsci.edu/cdbs/cdbs8/synphot\_tables/}. The filter transmission
curves for all systems are shown in Figs.\ \ref{figMultiPlot1} -
\ref{figColors}. The symbols represent the values given by {\tt K93b} or found
in the literature, the lines show the results of the (curvature weighted)
parabolic interpolations from the Kurucz programs, which are sometimes
convoluted with other response functions (atmosphere, optics, receiver).

\cp

\begin{table}[ht!]
 \begin{tabular}{|c|}
  \hline
  {\bfseries H-Beta} (Vega)\hfill\ \\
  \hline
  Kurucz 1993b, 1998 ({\tt .../programs/COLORS/})\\ 
  \hline
  \begin{tabular}{|c|}
    H$_\beta$ \\ \hline
     2.903    \\ \hline 
     2.695              
  \end{tabular}\\
  \hline
  Using additional response functions `OneP21', `Air', and `Reflct'. \\
  Output "468nm.dat" are not correct $H_\beta$ values. (see text below) \\
  \hline\hline 
  {\bfseries Delta a} (Vega)\hfill\ \\
  \hline
  Kupka et al.\ (2003),
  Crawford \& Barnes (1970), Maitzen \& Vogt (1983)\\
  \hline
  \begin{tabular}{|c|c|}
    $a$      & b$-$y    \\ \hline
    $-$0.009 & $-$0.005 \\ \hline 
    \hm0.600 & \hm0.500           
  \end{tabular}\\
  \hline
  $a=g_2-(g_1+y)/2$, $\Delta a = a-a_0[(b-y);(B-V);(g_1-y)]$\\
  Using additional response functions `OneP21' and `Reflct'. \\
  \hline\hline 
  {\bfseries DDO}\hfill\ \\
  \hline
  {\tt L82} p.\ 55\\
  \hline
  \begin{tabular}{|*{6}{c|}}
    35$-$38 & 38$-$41 & 41$-$42 & 42$-$45 & 45$-$48 \\ \hline
      -     & -       & -       & -       & -       \\ \hline 
     0.000  &  0.000  &  0.000  &  0.000  &  0.000            
  \end{tabular}\\
  \hline
  No zero point star given.\\
  Using additional response functions `OneP21', `Air', and `Reflct'.\\
  \hline\hline 
  {\bfseries uvbyCa} (Vega, {\itshape HD83373})\hfill\ \\
  \hline
  Crawford \& Barnes (1970), Hauck \& Mermilliod (1980),\\
  Str\"omgren (1966)\\
  \hline
  \begin{tabular}{|*{7}{c|}}
    u$-$b  & u$_0-$b & b$-$y & m$_1$    &    c$_1$ & c$_{10}$ &
                                              {\itshape hk}      \\ \hline
     1.411 &  1.411  & 0.004 & \hm0.157 & \hm1.089 & \hm1.089 &
                                              {\itshape\hm0.194} \\ \hline 
     0.802 &  0.780  & 0.500 & $-$0.070 & $-$0.058 & $-$0.080 &
                                              {\itshape $-$0.133}          
  \end{tabular}\\
  \hline
  Model for HD83373 (hk=0.194):\\
  (\teff, \logg, \mh, \vmic) = (9\,250, 4.0, +0.0, 2)\\
  Using additional response functions `OneP21', `Air', and `Reflct'.\\
  \hline       
 \end{tabular}
 \caption{Names, zero point stars, references, standard and correction values of
          indices of the photometric systems. The last line contains remarks
          concerning the Kurucz programs among others.}
 \label{tabBIG1}
\end{table}

It turned out that the standard output of \atlas\ for the flux files is meshed
too coarse in $\lambda$ to give accurate results for H$_\beta$. So the color
output for this system is called "468nm.dat" instead of "beta.dat".

\cp

\begin{table}[ht]
 \begin{tabular}{|c|}
  \hline
  {\bfseries Walraven} ($\beta$ Leo)\hfill\ \\
  \hline
  {\tt L82} p.\ 76, Lub \& Pel (1977)\\
  \hline
  \begin{tabular}{|*{4}{c|}}
      V$-$B  & B$-$U    & U$-$W    & B$-$L    \\ \hline
    \hm0.034 &  0.436   &  0.108   &  0.198   \\ \hline 
    $-$0.561 &  1.232   &  0.106   &  0.460           
  \end{tabular}\\
  \hline
  $\lambda=372\,$nm: $l_1({\tt K93b}) = 0.468$, $l_1({\tt L82})=0.488$\\
  $\lambda=462\,$nm: $b_1({\tt K93b}) = 0.282$, $b_1({\tt L82})=0.287$\\
  Model for $\beta$ Leo: (\teff, \logg, \mh, \vmic) = (8\,850, 4.16, +0.0, 2) \\
  \hline\hline 
  {\bfseries Geneva} (Vega)\hfill\ \\
  \hline
  {\tt L82} p.\ 61\\
  \hline
  \begin{tabular}{|*{7}{c|}}
    U$-$B & V$-$B & B$_1-$B  & B$_2-$B & V$_1-$B & G$-$B \\ \hline
    1.505 & 0.959 &  0.900   &  1.510  &  1.662  & 2.168 \\ \hline 
    0.759 & 0.293 &  0.942   &  1.443  &  1.045  & 1.318           
  \end{tabular}\\
  \hline
  {\tt K93b} has some additional response function values.\\
  \hline\hline 
  {\bfseries Geneva3} (Vega)\hfill\ \\
  \hline
  Cramer \& Maeder (1979), K\"unzli et al.\ (1997), {\tt L82} p.\ 61,\\
  North \& Kupka (1997), Rufener \& Nicolet (1988)\\
  \hline
  \begin{tabular}{|*{7}{c|}}
    U$-$B & V$-$B & B$_1-$B  & B$_2-$B & V$_1-$B & G$-$B \\ \hline
    1.505 & 0.959 &  0.900   &  1.510  &  1.662  & 2.168 \\ \hline 
    0.004 & 0.017 &  0.017   &  0.005  &  0.018  & 0.027           
  \end{tabular}\\
  \hline
  Additional (derived) indices are calculated:\\
  U$-$B$_2$, B$_1-$B$_2$, B$_2-$V$_1$, V$_1-$G, d, $\Delta$, g, m$_2$, X, Y, Z,
  pT, pG\\
  \hline\hline 
  {\bfseries UBV (USA $\mid$ Vilnius $\mid$ Buser)} (Vega)\hfill\ \\
  \hline
  Arp (1961),  A$\check{\rm z}$usienis \& Strai$\check{\rm z}$ys (1969),
    Buser (1978), \\ 
  Buser \& Kurucz (1978), {\tt L82} p.\ 48, Matthews \& Sandage (1963)\\
  \hline
  \begin{tabular}{|*{6}{c|}}
    \multicolumn{3}{c}{\vlx{U$-$B}}   \vline &
    \multicolumn{3}{c}     {B$-$V}    \vline \\ \hline
    \multicolumn{3}{c}{\vlx{$-$0.005}}\vline &
    \multicolumn{3}{c}     {$-$0.003} \vline \\ \hline          
     u$_1-$b$_1$       & u$_{1.3}-$b$_{1.3}$ & U$_3-$B$_2$      &
     b$_1-$v$_1$       &     b$_0-$v$_0$     & B$_3-$v$_0$      \\ \hline
     $-$0.475          & $-$0.468            & $-$0.408 &
        0.619          &    0.601            &    0.601         
  \end{tabular}\\
  \hline
  \\
  \hline\hline 
  {\bfseries Vilnius (7 color)} (Vega)\hfill\ \\
  \hline
  {\tt L82} p.\ 66\\
  \hline
  \begin{tabular}{|*{7}{c|}}
     U$-$P  & P$-$X & X$-$Y  & Y$-$Z  & Z$-$V  & V$-$S  & T$-$S  \\ \hline
      0.630 & 0.830 &  0.290 &  0.100 &  0.050 &  0.140 &  0.040 \\ \hline 
      0.304 & 0.302 &  0.543 &  0.414 &  0.218 &  0.826 &  0.267         
  \end{tabular}\\
  \hline
  Normalization in {\tt L82}: all indices are 0.000 for\\
  (\teff, \logg, \mh, \vmic) = (30\,000, 4.00, +0.0, 2)\\
  \hline       
 \end{tabular}
 \caption{see Table \ref{tabBIG1}}
 \label{tabBIG2}
\end{table}

\cp

\begin{table}[ht!]
 \begin{tabular}{|c|}
  \hline
  {\bfseries HST (`External')}\hfill\ \\
  \hline
  {\tt ftp://ftp.stsci.edu/cdbs/cdbs8/synphot\_tables/}\\
  \hline
  \begin{tabular}{|*{3}{c|}}
     555  &  606  &  814  \\ \hline
     -    &  -    &  -    \\ \hline 
    0.000 & 0.000 & 0.000           
  \end{tabular}\\
  \hline
  No zero point star.\\
  Response functions of cameras `pcf', `wf2f', `wf3f', `wf4f', and\\
  `wfpc2\_f' used (same color index for one wavelength region).\\
  \hline\hline 
  {\bfseries Thirteen} (Vega)\hfill\ \\
  \hline
  {\tt L82} p.\ 68, Johnson \& Mitchell (1975)\\
  \hline
  \begin{tabular}{|*{7}{c|}}
    33$-$52  &  35$-$52 & 37$-$52 & 40$-$52 & 45$-$52 & 52$-$58&52$-$58'\\\hline
    \hm0.052 & \hm0.035 &   0.043 &   0.013 &   0.001 & $-$0.008 & 0.001\\\hline
    $-$0.202 & $-$0.259 &   0.090 &   0.584 &   0.377 & \hm0.356 & 0.390\\\hline
  \end{tabular}\\
  \begin{tabular}{|*{6}{c|}}
     52$-$63 & 52$-$72 & 52$-$80 &  52$-$86 & 52$-$99&52$-$110 \\ \hline
    $-$0.011 &  0.008  &  0.010  & $-$0.004 &  0.011 &  0.000  \\ \hline
    \hm0.664 &  1.103  &  1.451  & \hm1.683 &  2.019 &  2.441   
  \end{tabular}\\
  \hline
  58'({\tt K93b}) = 58({\tt L82}), 58({\tt K93b}): no reference found;\\
  (52$-$58)$_{\rm Vega}$ = $-$0.008 ({\tt K93b}) vs.\ 0.001 ({\tt L82})\\
  \hline\hline 
  {\bfseries Cousins (Cape)} (Vega)\hfill\ \\
  \hline
  Cousins (1976), {\tt L82} p.\ 47, 79, 82\\
  \hline
  \begin{tabular}{|c|c|}
       V$-$R &    V$-$I \\ \hline
    $-$0.009 & $-$0.005 \\ \hline 
    \hm0.593 & \hm1.228           
  \end{tabular}\\
  \hline
  V({\tt K93b})=v$_0$(UBV$_{\rm Vilnius}$),\\
  R({\tt K93b})=r$_c$({\tt L82}),
  I({\tt K93b})=i$_c$({\tt L82}) (= 0.84 vs.\ 0.86 at $\lambda=840\,$nm),\\
  additional values for RCA31034 (`Cathode75'):\\
  $(\lambda,S(\lambda)) = (880,0.03), (890,0.01)$ \\
  \hline\hline 
  {\bfseries JHKcit} (Vega)\hfill\ \\
  \hline
  Kurucz 1993b, 1998 ({\tt .../programs/COLORS/})\\ 
  \hline
  \begin{tabular}{|*{3}{r|}}
    V$-$J     & V$-$H    & V$-$K    \\ \hline
        0.030 &  0.030   &  0.030   \\ \hline 
     19.181   & 16.444   & 14.226             
  \end{tabular}\\
  \hline
  Using additional response functions `atmos', `detec', and `block'. \\
  \hline\hline 
  {\bfseries (V)RIJKL} (Vega)\hfill\ \\
  \hline
  {\tt L82} p.\ 71\\
  \hline
  \begin{tabular}{|*{6}{r|}}
    V$-$R & V$-$I & V$-$J & V$-$K & V$-$L \\ \hline
     0.000 & 0.000 &  0.000 &  0.000 &  0.000 \\ \hline 
     0.717 & 1.488 &  2.651 &  4.850 &  6.699           
  \end{tabular}\\
  \hline
  {\tt K93b} uses L$^{II}$ of {\tt L82}.\\
  \hline       
 \end{tabular}
 \caption{see Table \ref{tabBIG1}}
 \label{tabBIG3}
\end{table}
 
\cp

\figureDSSN{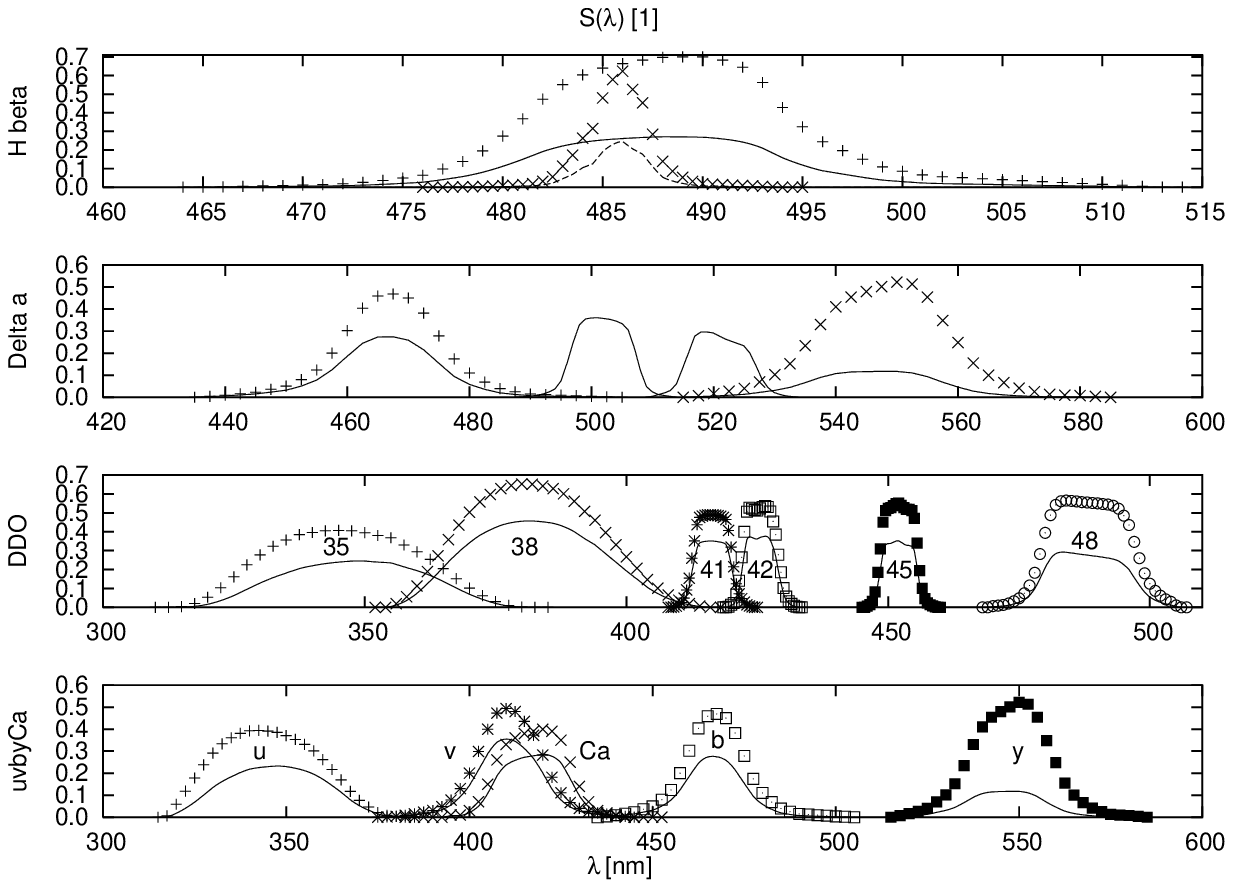}
{Response functions of four out of the 14 photometric systems (1/4). The symbols
 represent the filter values, the lines show the results of the convolution of
 their curvature weighted parabolic interpolations with other response functions
 (atmosphere: `Air' (not in $\Delta a$), optics: `Reflct', receiver: `OneP21').}
{figMultiPlot1}{ht!}{clip,angle=0,width=115mm}


\section*{Missing models - grid gaps}

The grid computations were performed with the perl package \smgt\ (Stellar
Model Grid Tool), described in Schmidt (1999, {\tt S99}). Due to bad initial
conditions or because the parameter region is physically and/or mathematically
`difficult', 1,110 out of the 91,520 models did not converge with respect to
the \smgt\ convergence criteria ({\tt S99}; {\tt H02}, their Table 3). Their
distribution in the grid is shown in Figures \ref{figMissT} - \ref{figMissC}
(p.~\pageref{figMissT} - \pageref{figMissC}). For the single parameters we find
difficult regions for low \teff\ values (Fig.~\ref{figMissT}), high (and low)
\logg\ values (Fig.~\ref{figMissG}), low \mh\ values (Fig.~\ref{figMissZ}), and
for CGM$_{288}$ whereas MLT$_{72}$ is `easiest' among the different convection
models (Fig.~\ref{figMissC}); \vmic\  seems not to be as relevant for creating
problems to achieve convergence (Fig.~\ref{figMissV}). A more detailed
insight---for combinations of two parameters---can be obtained from the 3D
plots of Figures \ref{figMissTG} - \ref{figMissGZ2}
(p.~\pageref{figMissTG} - \pageref{figMissGZ2}). More figures are available on
the \nemo\ website.

The total numbers of models fulfilling the primary and secondary \smgt\ 
convergence criteria are given in Table \ref{tabNumCrit}
(p.~\pageref{tabNumCrit}). In this sense the models are also called {\itshape
fully} and {\itshape partially converged}. Note that before interpolation (next
section) about 3/4 of the (90,410) models are fully converged and 1/4
partially,%

\figureDSSN{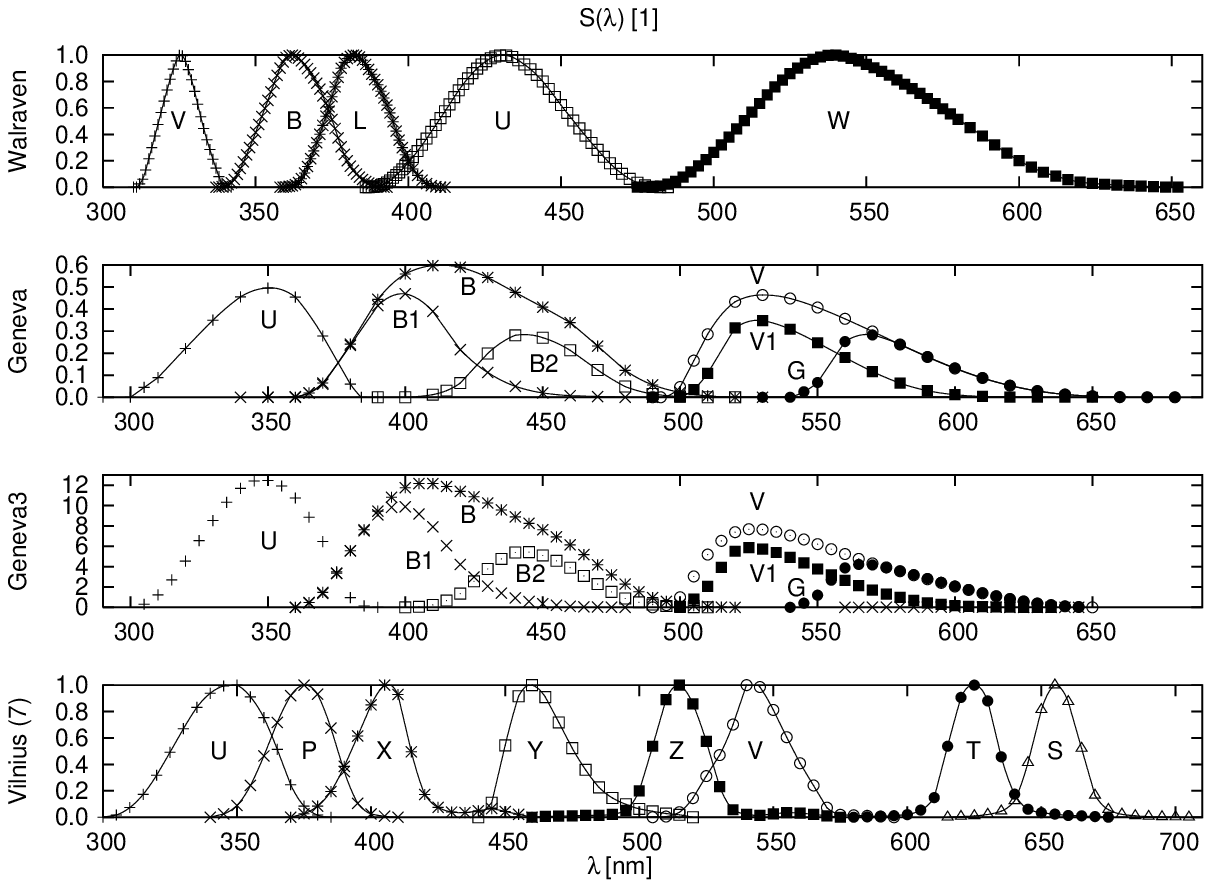} 
{Response functions (filter values and their curvature weighted parabolic
 interpolations) of four out of the 14 photometric systems (2/4).}
{figMultiPlot2}{t}{clip,angle=0,width=115mm}

\noindent
whereas with the (527) interpolated and consecutively converged files it is the
other way round: 1/4 fulfil the primary convergence criteria and 3/4 only the
secondary ones.


\section*{Interpolation}

\figureDSSN{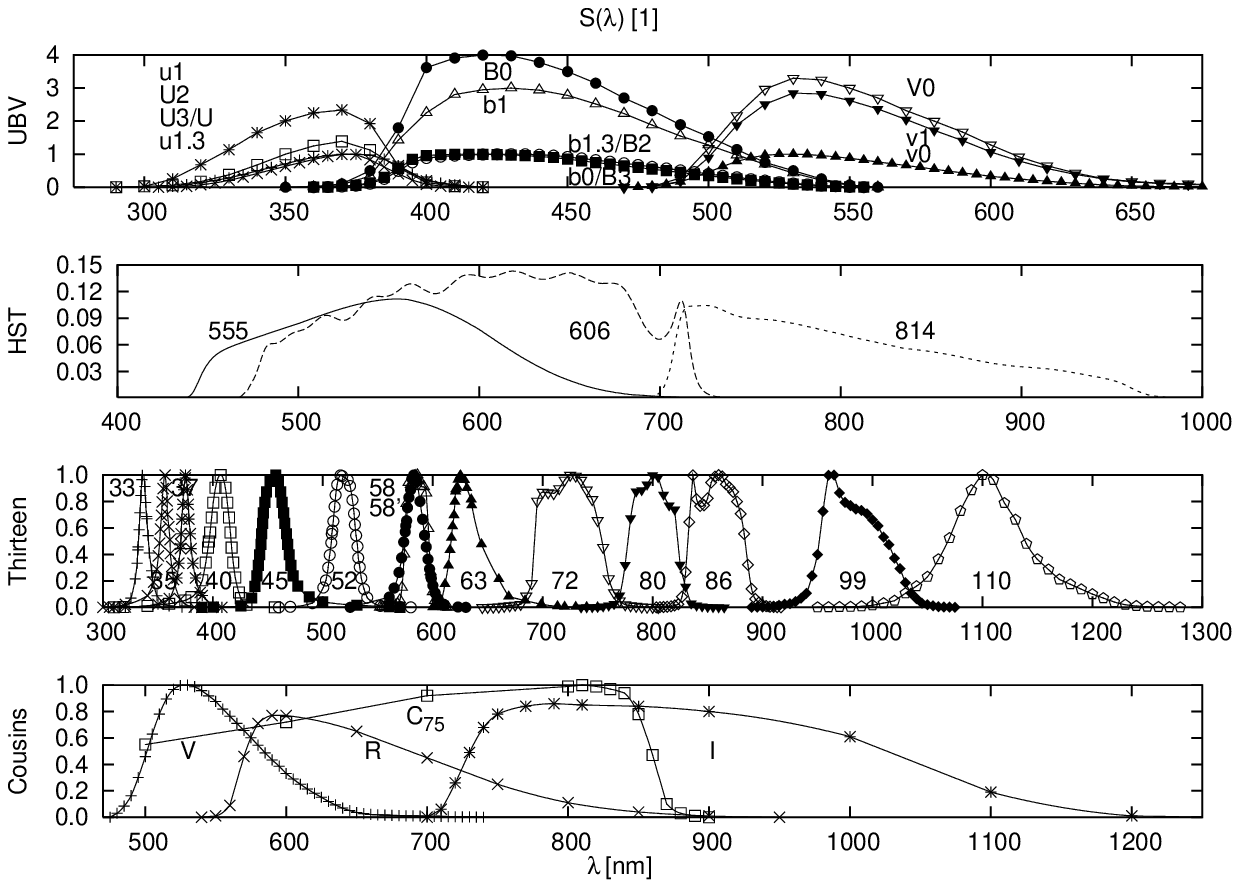}
{Response functions (filter values and their curvature weighted parabolic
 interpolations) of four out of the 14 photometric systems (3/4).}
{figMultiPlot3}{t}{clip,angle=0,width=115mm}

\figureDSSN{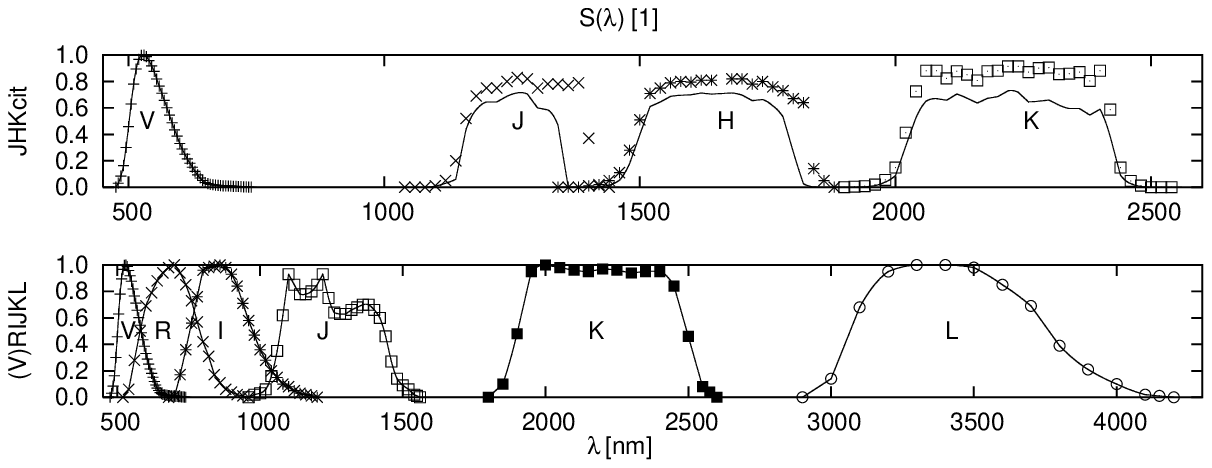}
{Response functions (filter values and their curvature weighted parabolic
 interpolations) of two out of the 14 photometric systems (4/4). JHKcit is
 convoluted with the additional response functions `atmos', `detec', and
 `block'.}
{figMultiPlot4}{h!}{clip,angle=0,width=115mm}

A {\tt Fortran90} program was written to close the gaps in the grid through
interpolation ({\tt Ip1}). Since an extended version of the program should
allow the interpolation of models (and their fluxes and colors) between the
grid nodes ({\tt Ip2}), the following requirements are desired:
\figureDSSN{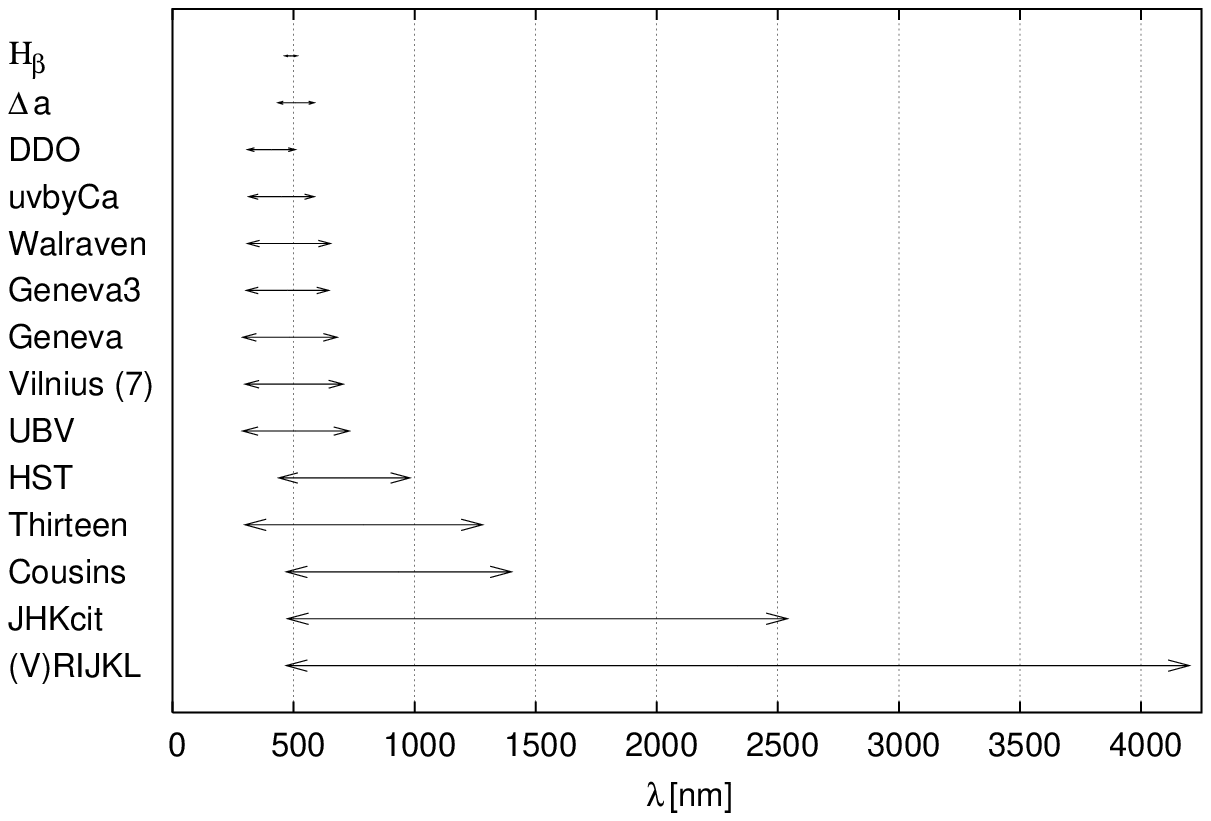}
{Wavelength range of the photometric systems}
{figColors}{t}{clip,angle=0,width=115mm}
\begin{enumerate}
  \item \label{itMult}
        Multidimensionality: As there are four numerical parameters---\teff,
        \logg, \mh, and \vmic\ (the convection model is non-numerical)---a 4D
        interpolation is necessary.

        This is realized by means of four 1D interpolations, one for each
        numerical parameter, because in the surroundings of a gap there are
        frequently further gaps. Hence, in general it is difficult to get a
        suitable 4D subgrid neighborhood.
  \item \label{itPrec}
        Precision: mmag in colors.

        Tests (comparisons of interpolated values with \smgt\ calculated ones)
        have shown that linear interpolation results in errors of some ten
        millimags, so at least parabolic 
        interpolation is necessary to shrink down the deviation to the desired
        accuracy of one millimag.
  \item \label{itLoc}
        Local procedure: A global procedure would involve 25,000 grid nodes and
        thus make the program complex and slow, and it requires a complete grid
        without gaps. The physics does not demand a global method either.

        This excludes splines which would satisfy the precision and
        differentiability requirements.
  \item \label{itDer} 
        Differentiability: First derivatives of color indices with respect to
        \teff\ and \logg\ are used for pulsation mode identification (Garrido et
        al.\ 1990); they should be computable and {\itshape continuous} at the
        grid nodes.

        Unlike curvature weighted parabolas as are used in the {\tt Fortran77}
        color programs by Kurucz ({\tt K93b}), {\itshape distance weighted
        parabolas} can provide continuous derivatives at the grid nodes and
        fulfill also the other requirements.
\end{enumerate}


\subsection*{Distance Weighted Parabolas for Interpolation}


\figureDSSN{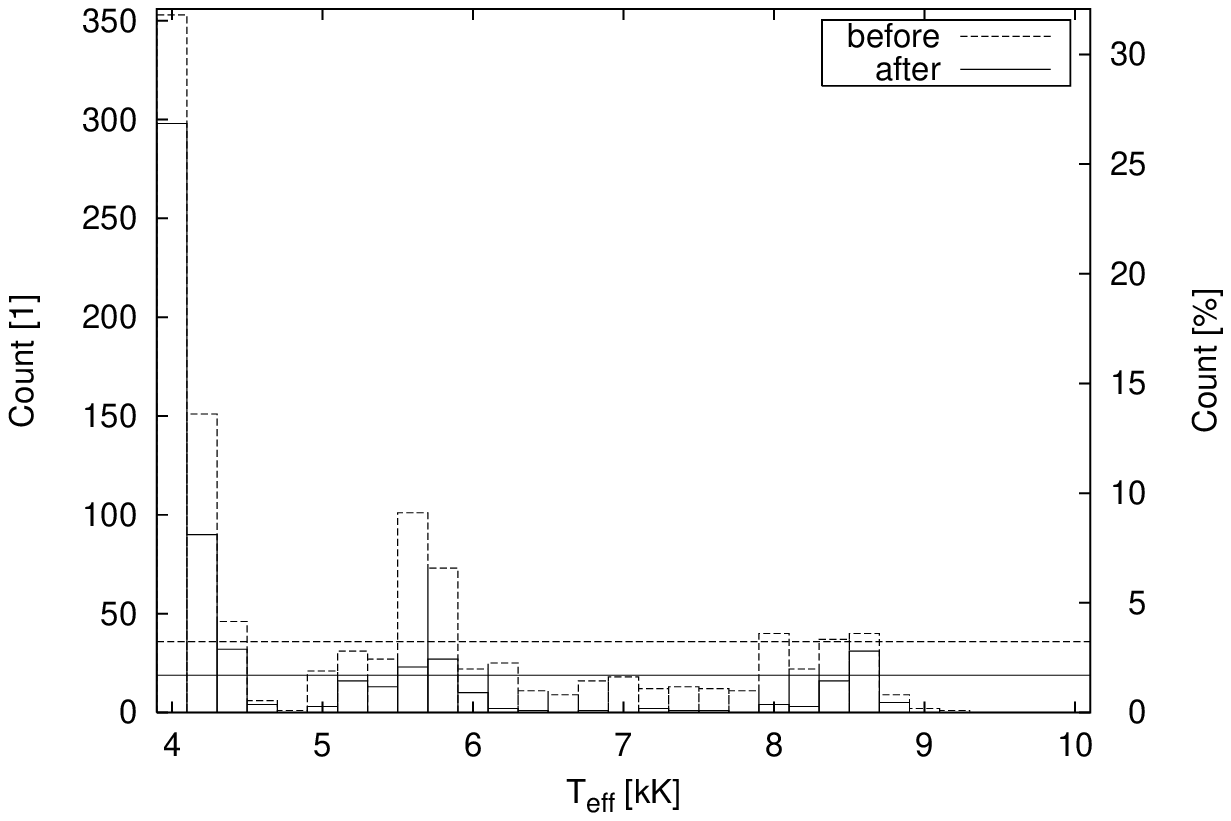}
{Number of missing models versus \teff\ (added up for the other four parameters)
 before and after the interpolation. The two horizontal lines indicate the
 average values per \teff\ step of 35.8 (=1,110/31) and 18.8 (=583/31).}
{figMissT}{t}{clip,angle=0,width=115mm}

Fig.~\ref{figIP} (p.~\pageref{figIP}) illustrates the interpolation procedure
in 1D. For an interpolation in the interval $x_2 \le x \le x_3$ we need two
left neighboring grid nodes $x_1$ and $x_2$ and two right ones $x_3$ and $x_4$.
An interpolating function $h(x)$ is constructed for the unknown function $f(x)$
using the four given  data points $(x_i,f_i), i=1,\dots,4$, where
\be f_i = f(x_i) \ee

We define a (quadratic) `left' parabola $p_{\rm L}(x)$ by means of
$x_{1,2,3}$ and a `right' one $p_{\rm R}(x)$ by means of $x_{2,3,4}$, and for
comparison a cubic polynomial $p_{\rm C}(x)$ by means of all four:
\bea
  \label{eqPx}
  p_{\rm L}(x_1) = & f_1 = p_{\rm C}(x_1) &                  \\
  p_{\rm L}(x_2) = & f_2 = p_{\rm C}(x_2) & = p_{\rm R}(x_2) \nn\\
  p_{\rm L}(x_3) = & f_3 = p_{\rm C}(x_3) & = p_{\rm R}(x_3) \nn\\
                   & f_4 = p_{\rm C}(x_4) & = p_{\rm R}(x_4) \nn
\eea

\figureDSSN{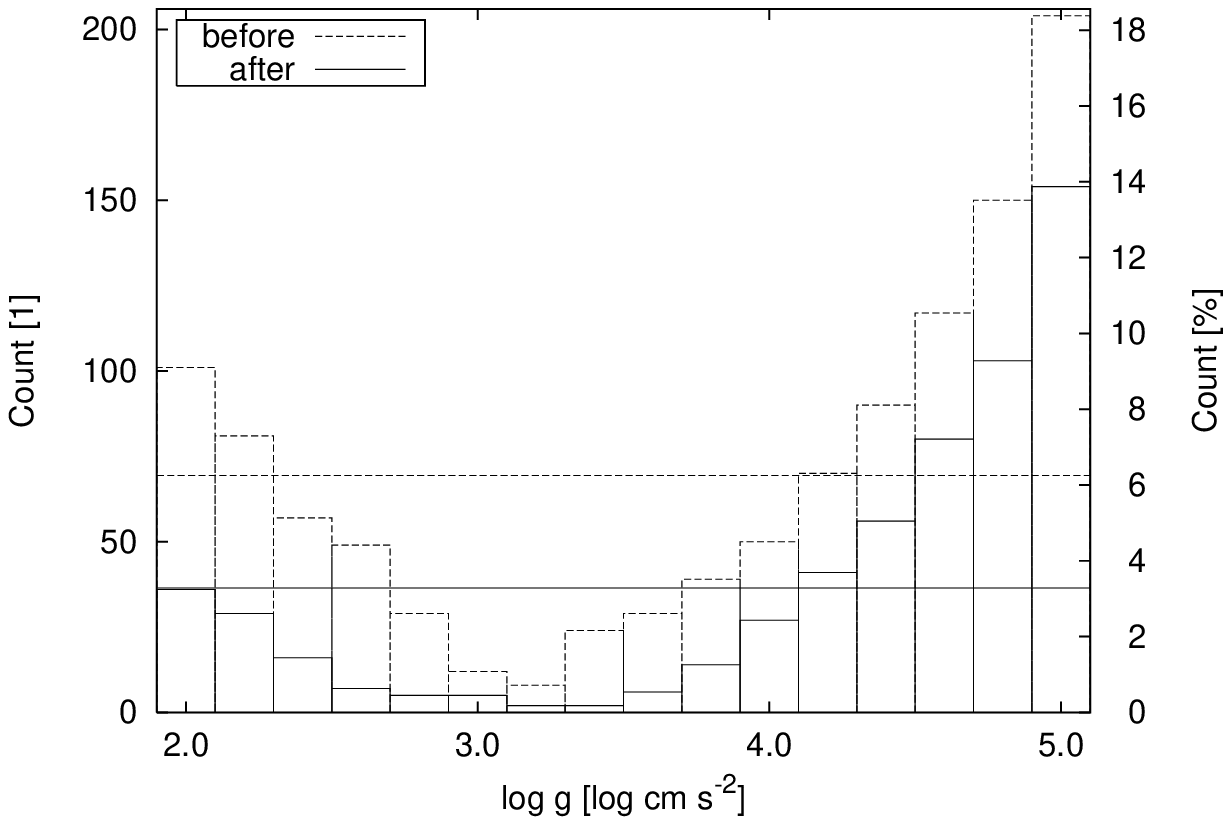}
{Number of missing models versus \logg\ (added up for the other four parameters)
 before and after the interpolation. The two horizontal lines indicate the
 average values per \logg\ step of 69.4 (=1,110/16) and 36.4 (=583/16).}
{figMissG}{t}{clip,angle=0,width=115mm}

The interpolating function for $x_2 \le x \le x_3$ is defined as distance
weighted mean of the two parabolas:
\bea
  \label{eqDefH}
  h(x) &=& \frac{w_{\rm L}(x) \cdot p_{\rm L}(x) +
                 w_{\rm R}(x) \cdot p_{\rm R}(x)}{w_{\rm L} + w_{\rm R}} = \\
       &=& \frac{(x_3-x)      \cdot p_{\rm L}(x) + 
                 (x-x_2)      \cdot p_{\rm R}(x)}{x_3-x_2} = \nn\\
       &=& \frac{\Delta_{30}  \cdot p_{\rm L}(x) + 
                 \Delta_{02}  \cdot p_{\rm R}(x)}{\Delta_{32}} \nn
\eea
which is a polynomial of order 3 in $x$. Note, that this method for
interpolation is not restricted to equidistant grids and that $w_{\rm L} +
w_{\rm R}$ is constant with respect to $x$. (The definition of $\Delta_{ij}$
is given in Eq.~\ref{eqDelta} on page \pageref{eqDelta}.)

It holds by (\ref{eqPx}) that
\bea
  h(x_2) \hspace{3mm}  = p_{\rm L}(x_2) &=&
    f_2  \hspace{5mm} (= p_{\rm R}(x_2) = p_{\rm C}(x_2) )\\
  h(x_3) \hspace{3mm}  = p_{\rm R}(x_3) &=&
    f_3  \hspace{5mm} (= p_{\rm L}(x_3) = p_{\rm C}(x_3) )\nn
\eea

\figureDSSN{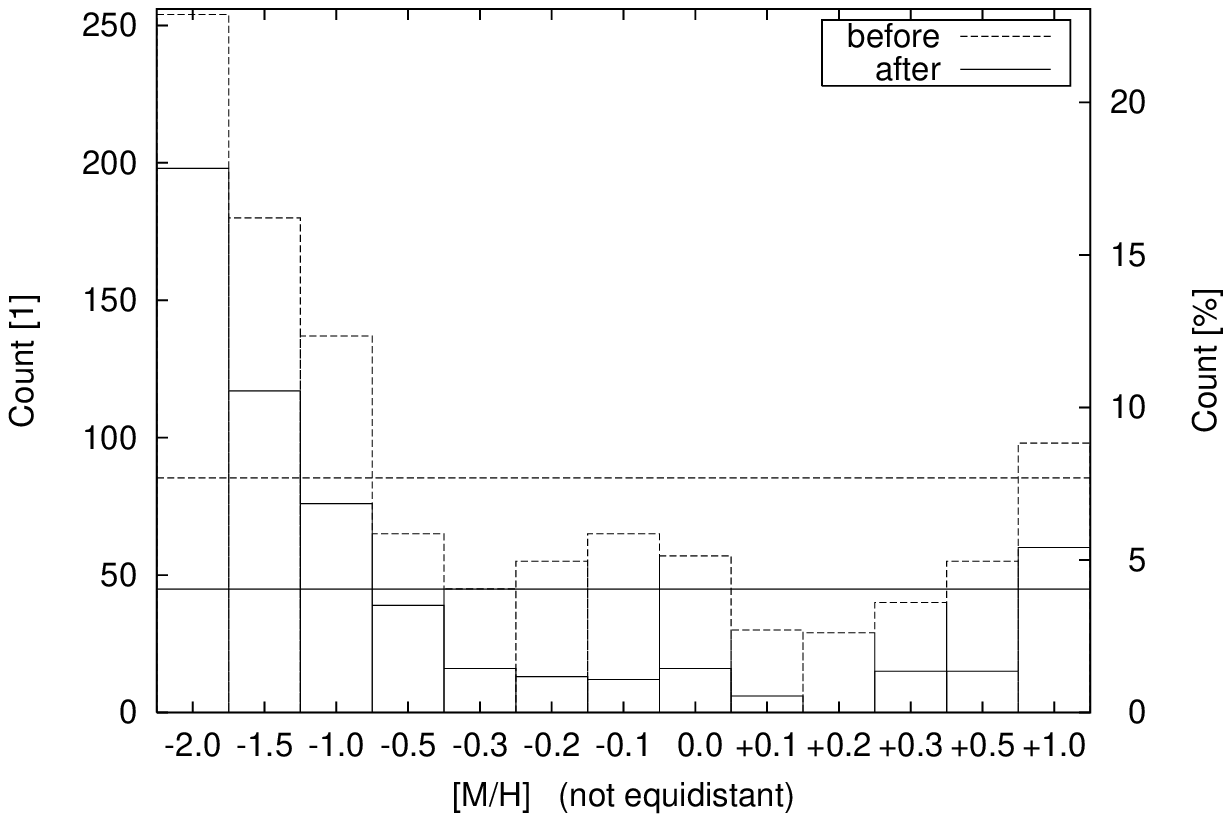}
{Number of missing models versus \mh\ (added up for the other four parameters)
 before and after the interpolation. The two horizontal lines indicate the
 average values per \mh\ step of 85.4 (=1,110/13) and 44.8 (=583/13).}
{figMissZ}{t}{clip,angle=0,width=115mm}

For the first derivative we obtain
\bea
  \label{eqDerH}
  h'(x)
  &=& \frac{-1\cdot p_{\rm L}(x) + (x_3-x)\cdot p'_{\rm L}(x) + \nn
             1\cdot p_{\rm R}(x) + (x-x_2)\cdot p'_{\rm R}(x)}{x_3-x_2} =\\
  &=& \frac{w_{\rm L}(x) \cdot p'_{\rm L}(x) + 
            w_{\rm R}(x) \cdot p'_{\rm R}(x)}{w_{\rm L} + w_{\rm R}} +
      \frac{[p_{\rm R} - p_{\rm L}](x)}{x_3-x_2}
\eea
which can be interpreted as the sum of the distance weighted derivatives and
the interval averaged `slope' between the right and left parabolas. The second
term vanishes for $x=x_2$ and $x=x_3$ and hence
\bea
  h'(x_2) &=& p'_{\rm L}(x_2)\\
  h'(x_3) &=& p'_{\rm R}(x_3)\nn
\eea

Analogously, if we define left and right parabolas $\tilde p_{\rm L}$($x$;
$x_{2,3,4}$) and $\tilde p_{\rm R}$($x$; $x_{3,4,5}$) (with $\tilde p_{\sf
L}(x)=p_{\rm R}(x)$ because of the same three underlying grid nodes $x_2$,
$x_3$, and $x_4$), and the interpolating function $h(x)$ as the distance
weighted mean of the two parabolas for the interval $x_3 \le x \le x_4$, we
obtain
\bea
  \label{eqConDer} 
  h'(x_3) &=& \tilde p'_{\rm L}(x_3) = p'_{\rm R}(x_3)\\
  h'(x_4) &=& \tilde p'_{\rm R}(x_4)\nn
\eea
\figureDSSN{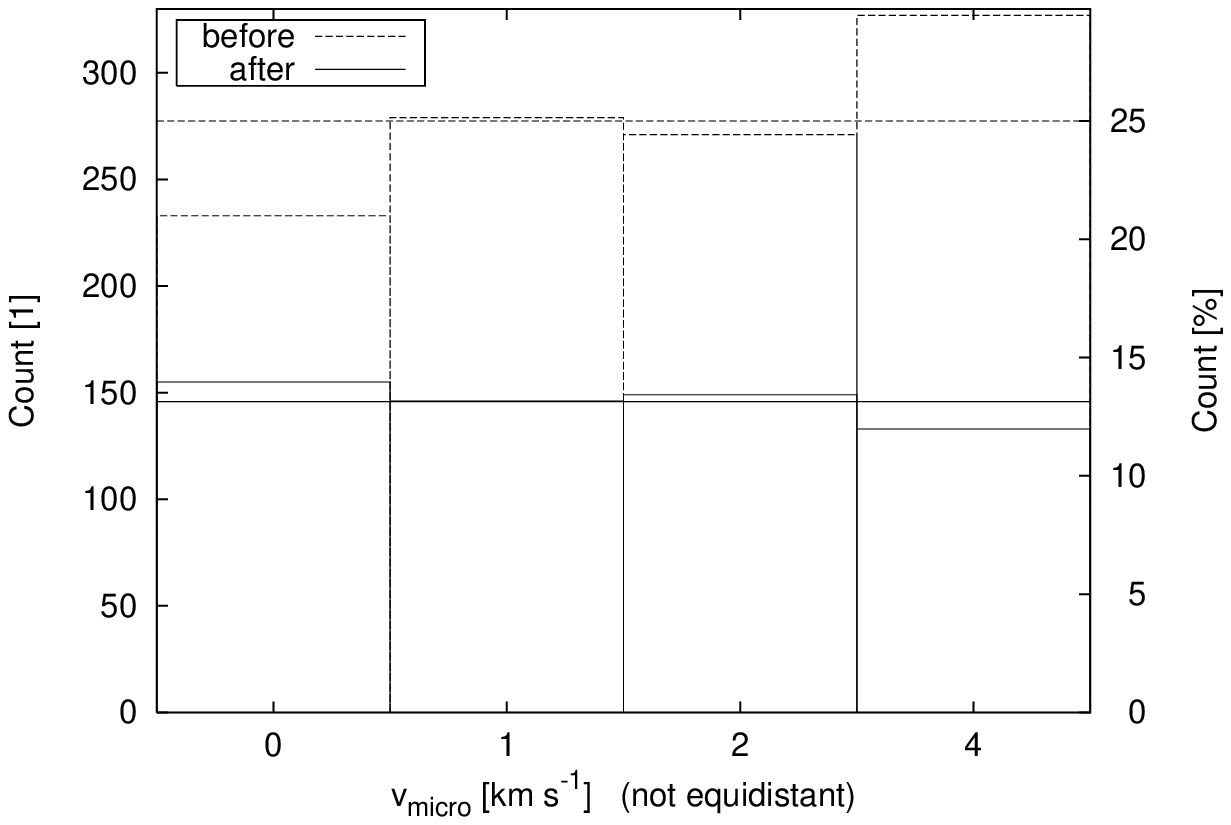}
{Number of missing models versus \vmic\ (added up for the other four parameters)
 before and after the interpolation. The two horizontal lines indicate the
 average values per \vmic\ step of 278 (=1,110/4) and 146 (=583/4).}
{figMissV}{t}{clip,angle=0,width=115mm}%
Eq.~\ref{eqConDer} shows the continuity of the first derivative of the
interpolating function $h'(x)$ at the grid nodes, and as (\ref{eqDerH}) yields a
uniquely defined quadratic polynomial in-between grid nodes, the continuity of
the first derivative as obtained from distance weighted parabolic interpolation
(\ref{eqPx})-(\ref{eqDefH}) holds for the entire of $x_1 \le x \le x_N$ for
which interpolation is required. Due to this property and the fact of being
third order in $x$, (\ref{eqPx})-(\ref{eqDefH}) is sometimes also called a
``local spline interpolation".

Using the definitions\footnote{We will never use it in the following ambiguous
manner: $x_0-x_j = \Delta_{0j} \neq \Delta_{0j}(x_i) = x_i-x_j$, except for the
unique case $i=0$.}
\bea
  \label{eqDelta}
  \Delta_{0j}(x)            &=& x  -x_j \\
  \Delta_{ij}\hspace{5.2mm} &=& x_i-x_j = -\Delta_{ji}
                                        =  \Delta_{ik} + \Delta_{kj}
                                        =  \Delta_{0j}(x_i) \nn
\eea
\figureDSSN{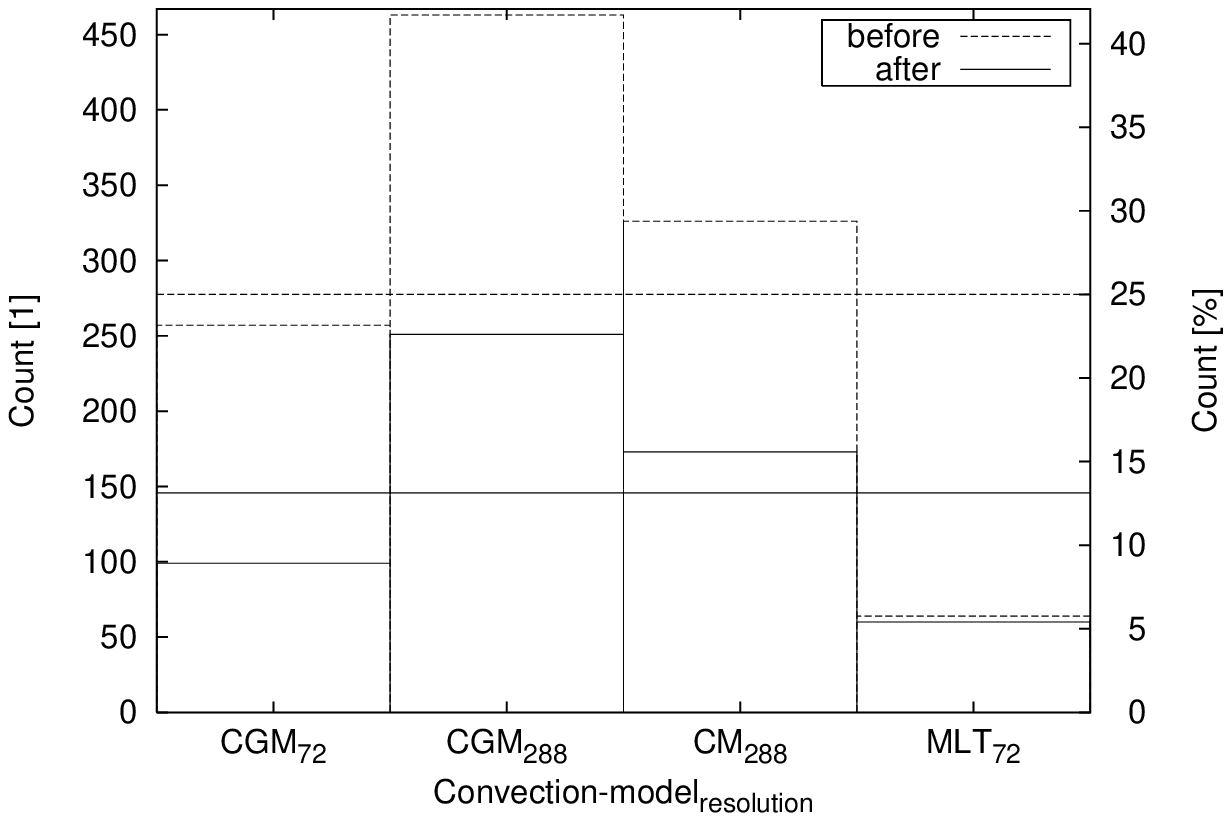}
{Number of missing models versus convection model (added up for the other four
 parameters) before and after the interpolation. The two horizontal lines
 indicate the average values per convection model of 278 (=1,110/4) and 146
 (=583/4).}
{figMissC}{t}{clip,angle=0,width=115mm} 
and Lagrange polynomials ($i,j,k=1,\dots,3$ or $2,\dots,4$),\footnote{$
  L_{ijkl}(x) = (\Delta_{0j}/\Delta_{ij}) \cdot (\Delta_{0k}/\Delta_{ik}) \cdot
                (\Delta_{0l}/\Delta_{il}), \\ 
  p_{\rm C}(x) = f_1 L_{1234}(x) + f_2 L_{2341}(x) + 
                 f_3 L_{3412}(x) + f_4 L_{4123}(x)$}
\bea
  \label{eqLP}
  L_{ijk}(x)   &=& \ffDelt{0j}{ij}{0k}{ik} 
                   = \frac{x-x_j}{x_i-x_j}\frac{x-x_k}{x_i-x_k}
                   = L_{ikj}(x) \\
  L_{ijk}(x_l) &=& \delta_{il}
                =  \left\{ {1,\ l=i \atop 0,\ l \neq i} \right.,\ 
                   l \in \{i,j,k\} \nn\\
  p_{\rm L}(x) &=& f_1 L_{123}(x) + 
                   f_2 L_{231}(x) + 
                   f_3 L_{312}(x) \\
  p_{\rm R}(x) &=& \hspx{$f_1 L_{123}(x) + {}$} 
                   f_2 L_{234}(x) + 
                   f_3 L_{342}(x) + 
                   f_4 L_{423}(x) \nn
\eea

\figureDSSN{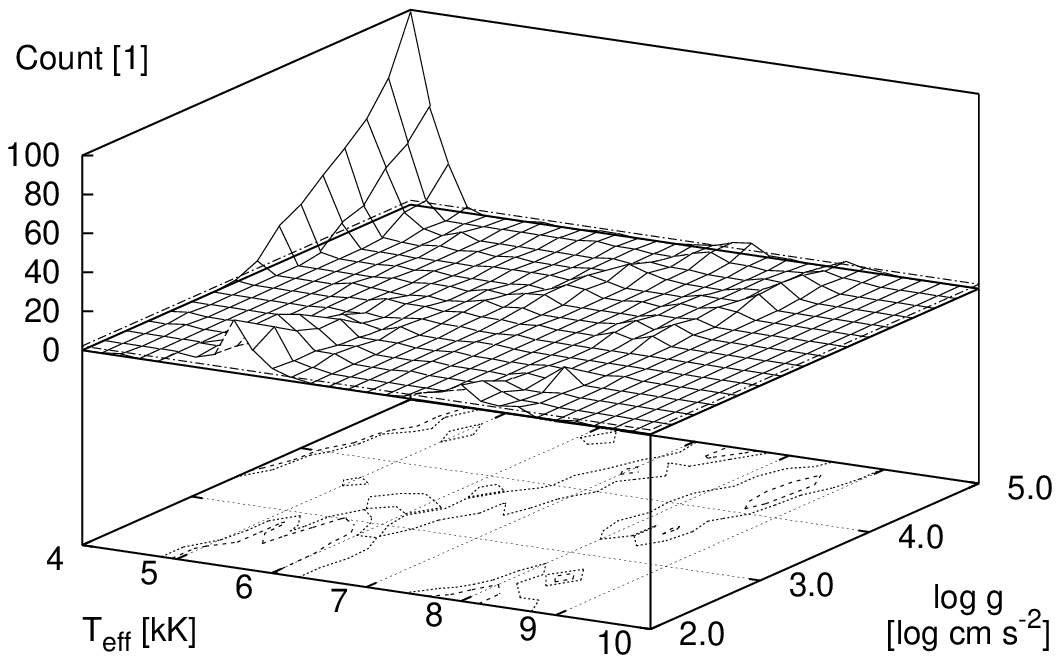}
{Number of missing models versus \teff\ and \logg\ (added up for the other three
 parameters) {\bfseries before} the interpolation. The contour lines are drawn
 for the half and the double of the average value of 2.24 (=1,110/31/16) which
 itself is indicated by the bordering dashed line in the 3D part of the plot.}
{figMissTG}{t}{clip,angle=0,width=115mm}
\noindent
we obtain
\bea
  \label{eqIp1D}
  \vec h(x) &=&
    \frac1{\Delta_{32}} \left\{
    \vec f_1 \right. \ub{       w_{\rm L}(x) L_{123}(x)}_{\bar L_1(x)} + 
    \vec f_4         \ub{       w_{\rm R}(x) L_{423}(x)}_{\bar L_4(x)} + {} \\
  && {} +
    \vec f_2         \ub{\left[ w_{\rm L}(x) L_{231}(x) +
                                w_{\rm R}(x) L_{234}(x) \right]}_{\bar L_2(x)} +
     {} \nn\\
  && {} +
    \vec f_3         \ub{\left[ w_{\rm L}(x) L_{312}(x) +
                                w_{\rm R}(x) L_{342}(x) \right]}_{\bar L_3(x)}
    \left. \rule{0mm}{3.3mm} \right\} = \nn\\
  &=&
  \frac1{\Delta_{32}} \left\{
    \frac1{\Delta_{32}} \left[
      \vec f_2\Delta_{30} \left(
        \frac{\Delta_{30}\Delta_{01}}{\Delta_{21}} +
        \frac{\Delta_{40}\Delta_{02}}{\Delta_{42}}\right) + \right. \right. \\
  & & \makebox[15.4mm]{} \left. +
      \vec f_3\Delta_{02}
        \left(\frac{\Delta_{30}\Delta_{01}}{\Delta_{31}} +
              \frac{\Delta_{40}\Delta_{02}}{\Delta_{43}}\right) \right] - \nn\\
  & & \makebox[9.5mm]{} \left. - \nn 
    \Delta_{30}\Delta_{02} \left[
      \vec f_1\frac{\Delta_{30}}{\Delta_{31}\Delta_{21}} +
      \vec f_4\frac{\Delta_{02}}{\Delta_{43}\Delta_{42}} \right] \right\}
\eea

\noindent
where we have taken into account that we also want to interpolate vector
functions in the model grid, because every $\vec f_i$ represents a model
atmosphere 
with up to $6\cdot 288=1728$ numerical values. Therefore, the formula is
optimized for computational performance\footnote{{\tt Fortran90} array
arithmetic} such that each $\vec f_i$ occurs only once (and all $\Delta_{ij}
\ge 0$ for $x_1 < x_2 \le x=x_0 \le x_3 < x_4$).

\figureDSSN{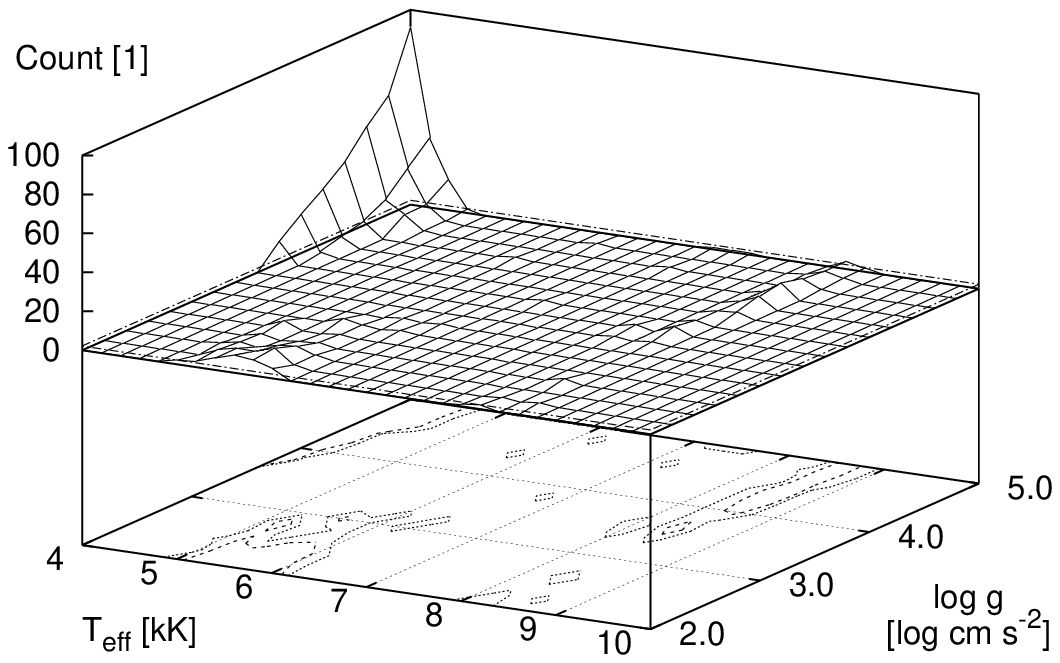}
{Number of missing models versus \teff\ and \logg\ (added up for the other three
 parameters) {\bfseries after} the interpolation. The contour lines are drawn
 for the half and the double of the average value of 1.18 (=583/31/16) which
 itself is indicated by the bordering dashed line in the 3D part of the plot.}
{figMissTG2}{t}{clip,angle=0,width=115mm}


\subsubsection*{Interpolation at grid nodes ({\tt Ip1})}

Formula (\ref{eqIp1D}) is used for closing the gaps in the model atmosphere
grids. Assume a missing model at
\be
  \label{eqMisMod}
  \vec x_0 = (t_0,g_0,z_0,v_0[,c_0])
\ee
where
\bea
      t & = x_{(1)} & = \teff      \\
      g & = x_{(2)} & = \log g \nn \\
      z & = x_{(3)} & = \mh    \nn \\
      v & = x_{(4)} & = \vmic  \nn \\
 \ [\ c & = x_{(5)} & = {\sf Convection\ model\ _{resolution}}\ ] \nn 
\eea

\figureDSSN{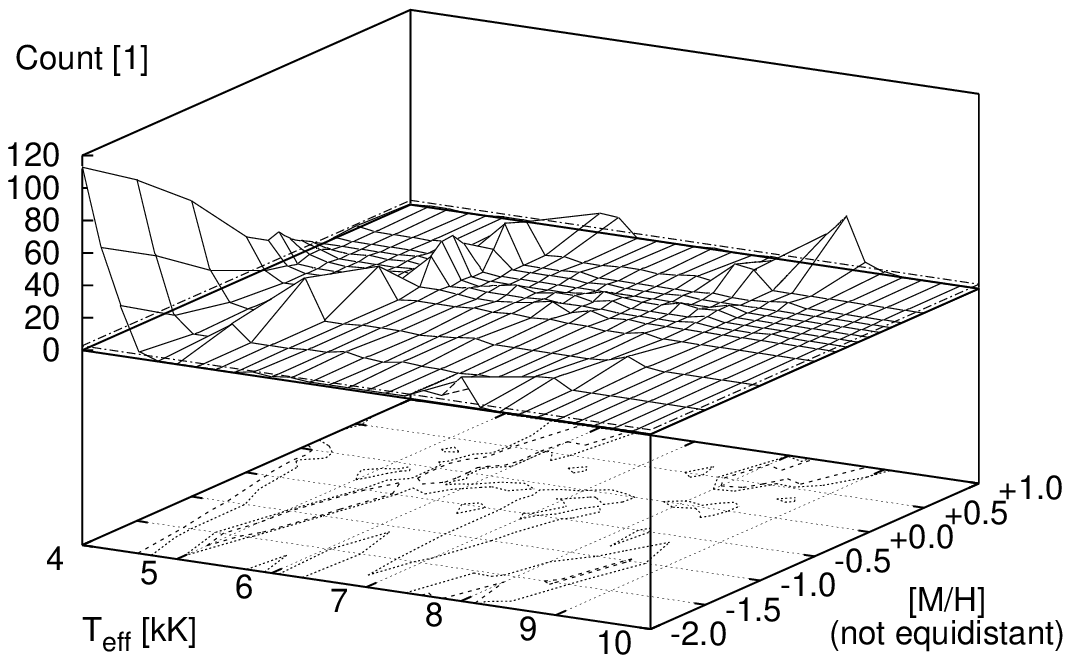}
{Number of missing models versus \teff\ and \mh\ (added up for the other three
 parameters) {\bfseries before} the interpolation. The contour lines are drawn
 for the half and the double of the average value of 2.75 (=1,110/31/13) which
 itself is indicated by the bordering dashed line in the 3D part of the plot.}
{figMissTZ}{t}{clip,angle=0,width=115mm}

Note, that $c=x_{(5)}$ is a non-numerical parameter of the grid and hence can
not be used for interpolation. Therefore, we have to consider four 4D subgrids
(for each value of $c$) and neglect $c$ hereafter when talking about
interpolation.

For each $x_{(s)}$, $s=1,\dots,4$, an independent 1D interpolation is done in
the following manner:

First we look for the next two left and right neighbors
in the grid. Their parameters and the corresponding (vector) function
values\footnote{In the case of \atlas model atmospheres the quantities
representing the atmospheric structure are the integrated mass, the
temperature, the gas pressure, the electron density, the Rosseland mean opacity
and the radiative acceleration at each depth.} are:

for $s=1$: $x_{(s)}=x_{(1)}=t$ ($t_1 < t_2 \le t_0 \le t_3 < t_4$):
\bea
  {\sf left}    &:& \vec x_{1(1)} = (t_1,g_0,z_0,v_0), \ 
                    \vec x_{2(1)} = (t_2,g_0,z_0,v_0) \\
  {\sf right}   &:& \vec x_{3(1)} = (t_3,g_0,z_0,v_0), \ \nn
                    \vec x_{4(1)} = (t_4,g_0,z_0,v_0)  \\
  \vec f_{i(1)} &=& \vec f(\vec x_{i(1)}) = \vec f(t_i,g_0,z_0,v_0) \nn
\eea

\figureDSSN{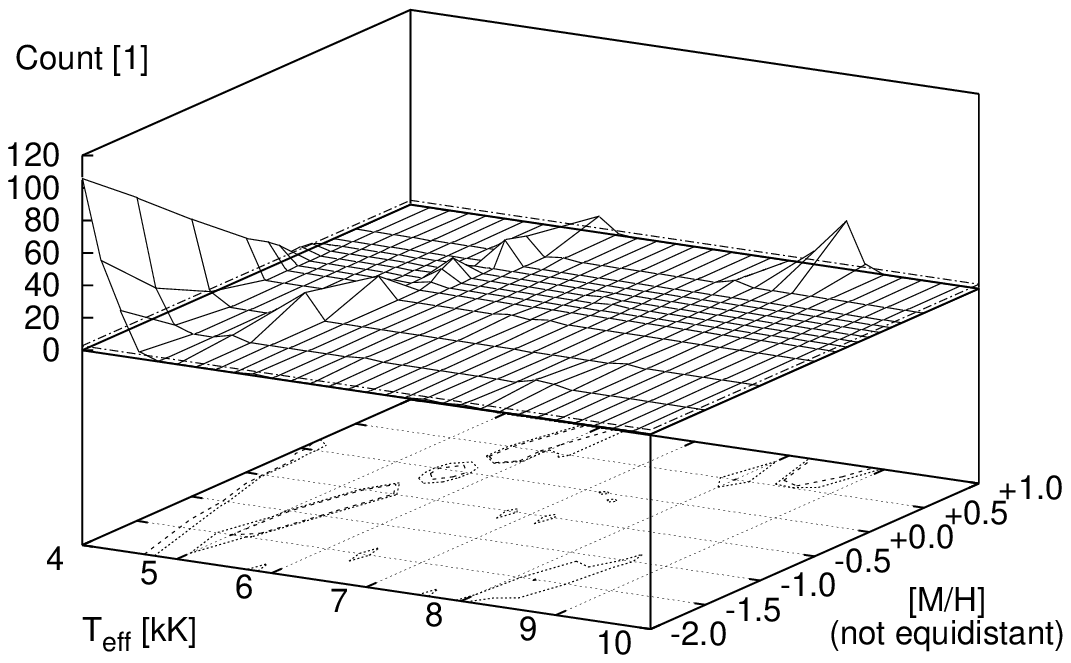}
{Number of missing models versus \teff\ and \mh\ (added up for the other three
 parameters) {\bfseries after} the interpolation. The contour lines are drawn
 for the half and the double of the average value of 1.45 (=583/31/13) which
 itself is indicated by the bordering dashed line in the 3D part of the plot.}
{figMissTZ2}{t}{clip,angle=0,width=115mm}

for $s=2$: $x_{(s)}=x_{(2)}=g$ ($g_1 < g_2 \le g_0 \le g_3 < g_4$):
\bea
  {\sf left}    &:& \vec x_{1(2)} = (t_0,g_1,z_0,v_0), \ 
                    \vec x_{2(2)} = (t_0,g_2,z_0,v_0) \\
  {\sf right}   &:& \vec x_{3(2)} = (t_0,g_3,z_0,v_0), \ \nn
                    \vec x_{4(2)} = (t_0,g_4,z_0,v_0)  \\
  \vec f_{i(2)} &=& \vec f(\vec x_{i(2)}) = \vec f(t_0,g_i,z_0,v_0) \nn
\eea

for $s=3$: $x_{(s)}=x_{(3)}=z$ ($z_1 < z_2 \le z_0 \le z_3 < z_4$):
\bea
  {\sf left}    &:& \vec x_{1(3)} = (t_0,g_0,z_1,v_0), \ 
                    \vec x_{2(3)} = (t_0,g_0,z_2,v_0) \\
  {\sf right}   &:& \vec x_{3(3)} = (t_0,g_0,z_3,v_0), \ \nn
                    \vec x_{4(3)} = (t_0,g_0,z_4,v_0)  \\
  \vec f_{i(3)} &=& \vec f(\vec x_{i(3)}) = \vec f(t_0,g_0,z_i,v_0) \nn
\eea

\figureDSSN{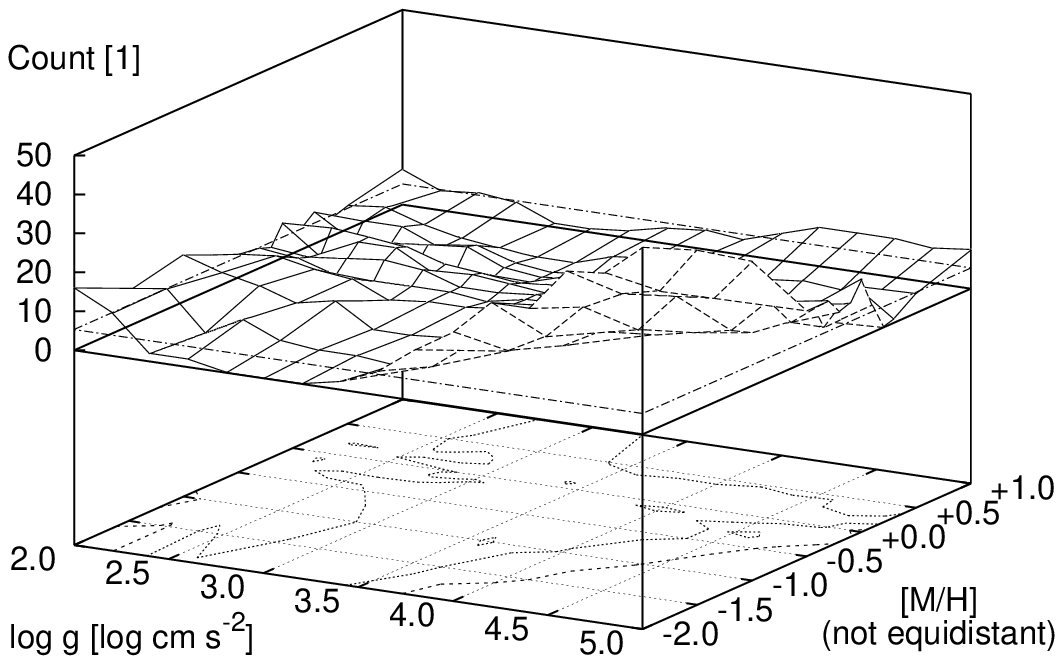}
{Number of missing models versus \logg\ and \mh\ (added up for the other three
 parameters) {\bfseries before} the interpolation. The contour lines are drawn
 for the half and the double of the average value of 5.34 (=1,110/16/13) which
 itself is indicated by the bordering dashed line in the 3D part of the plot.}
{figMissGZ}{t}{clip,angle=0,width=115mm}

for $s=4$: $x_{(s)}=x_{(4)}=v$ ($v_1 < v_2 \le v_0 \le v_3 < v_4$):
\bea
  {\sf left}    &:& \vec x_{1(4)} = (t_0,g_0,z_0,v_1), \ 
                    \vec x_{2(4)} = (t_0,g_0,z_0,v_2) \\
  {\sf right}   &:& \vec x_{3(4)} = (t_0,g_0,z_0,v_3), \ \nn
                    \vec x_{4(4)} = (t_0,g_0,z_0,v_4)  \\
  \vec f_{i(4)} &=& \vec f(\vec x_{i(4)}) = \vec f(t_0,g_0,z_0,v_i) \nn
\eea
Subsuming the third lines we can write\footnote{One remark about notation: 
$\vec x   = (x_{(1)}, x_{(2)},x_{(3)},x_{(4)}[,x_{(5)}]) = (t,g,z,v[,c])$,
$\vec x_0 = (x_{(1)0}, x_{(2)0},x_{(3)0},x_{(4)0}[,x_{(5)0}]) =
            (t_0,g_0,z_0,v_0[,c_0])$;
e.g.\ s=3:
  {\boldmath$\vec x_{i(3)}$}$ = (t_0,g_0,${\boldmath$z_i$}$,v_0[,c_0]) =
  (x_{(1)0},x_{(2)0},${\boldmath$x_{(3)i}$}$,x_{(4)0}[,x_{(5)0}])$.
I.e.\ $x_{(s)i}$ is a component of $\vec x_{i(s)}$. Thus, we could define 
$m_{(s)}(\vec x_{i(s)}) = m_{x_{(s)i}}$ (Eqs.~\ref{eqWeightFac},
\ref{eqGeneral}).}
\be
  \vec f_{i(s)} = \vec f(\vec x_{i(s)})
\ee

With Eq.~\ref{eqIp1D} (p.~\pageref{eqIp1D},
$\Delta_{ij}\rightarrow\Delta_{ij(s)}$) the 1D interpolation for $x_{(s)}$ at
$\vec x_0$ can be done:
\be
  \vec h_{(s)}(\vec x_0) = \vec h_{(s)}(t_0,g_0,z_0,v_0) = 
  \frac1{\Delta_{32(s)}} \sum_{i=1}^4 \vec f_{i(s)} \bar L_i(x_{(s)0})
\ee

\figureDSSN{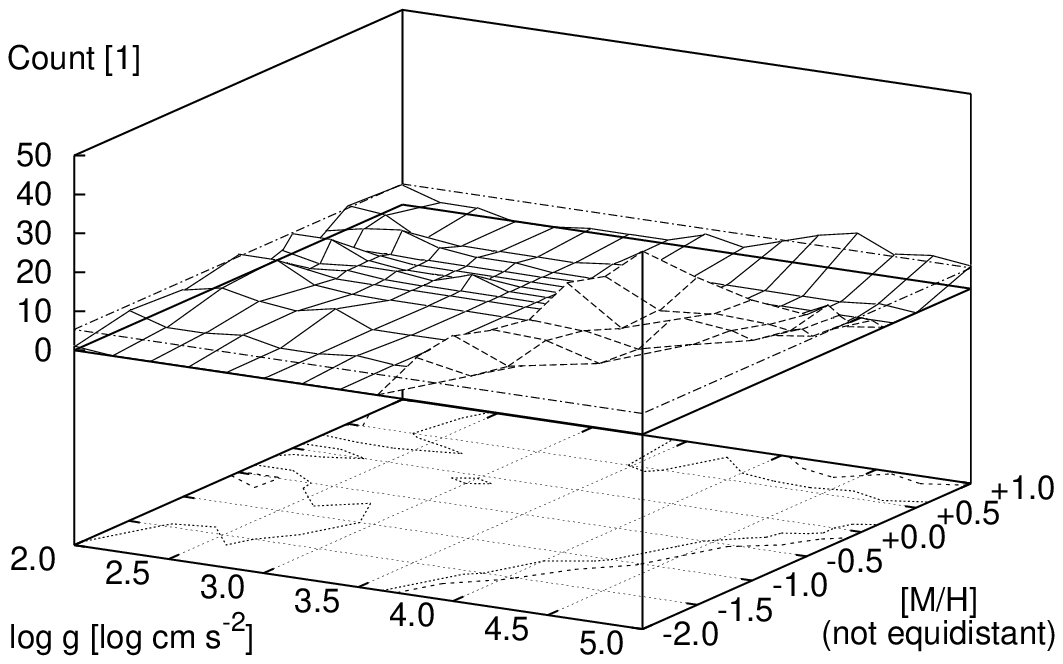}
{Number of missing models versus \logg\ and \mh\ (added up for the other three
 parameters) {\bfseries after} the interpolation. The contour lines are drawn
 for the half and the double of the average value of 2.80 (=583/16/13) which
 itself is indicated by the bordering dashed line in the 3D part of the plot.}
{figMissGZ2}{t}{clip,angle=0,width=115mm}

We introduce weighting factors to allow for neighbors which are not at the
next grid positions because there are further gaps in between:
\bea
  \label{eqWeightFac}
  w_{(s)} &=& \frac{6}{d_{(s)}} \\
  d_{(s)} &=& \sum_{i=1}^4|m_{x_{(s)0}}-m_{x_{(s)i}}| \nn
\eea
where $m_{x_{(s)i}}$ is the grid position number of $x_{(s)i}$ (number
corresponding to $x_{(s)i}$ when counting every grid step from the minimum to
the maximum value of parameter $x_{(s)}$).\footnote{E.g.\ for $s=1$, $x_{(s)i}
= x_{(1)i} = t_i = T_{{\rm eff},i} = 4\,000$: $m_{x_{(s)i}} = 1$ ($2$ for
$4\,200$, $3$ for $4\,400$, \dots, $31$ for $10\,000$). Note, that we are
considering interpolation at grid nodes (missing models) and consequently
$m_{x_{(s)0}}$ is also defined uniquely.}\\
Thus, $d_{(s)}$ is the sum of distances in grid step units of the interpolating
point $x_{(s)0}$ from the neighbors $x_{(s)i}$ (or in point of view of 4D: sum
of distances of $\vec x_{0}$ from $\vec x_{i(s)}$). In the optimal case it is
$2+1+1+2=6$ and the maximum weighting factor is $1.0$.

\figureDSSN{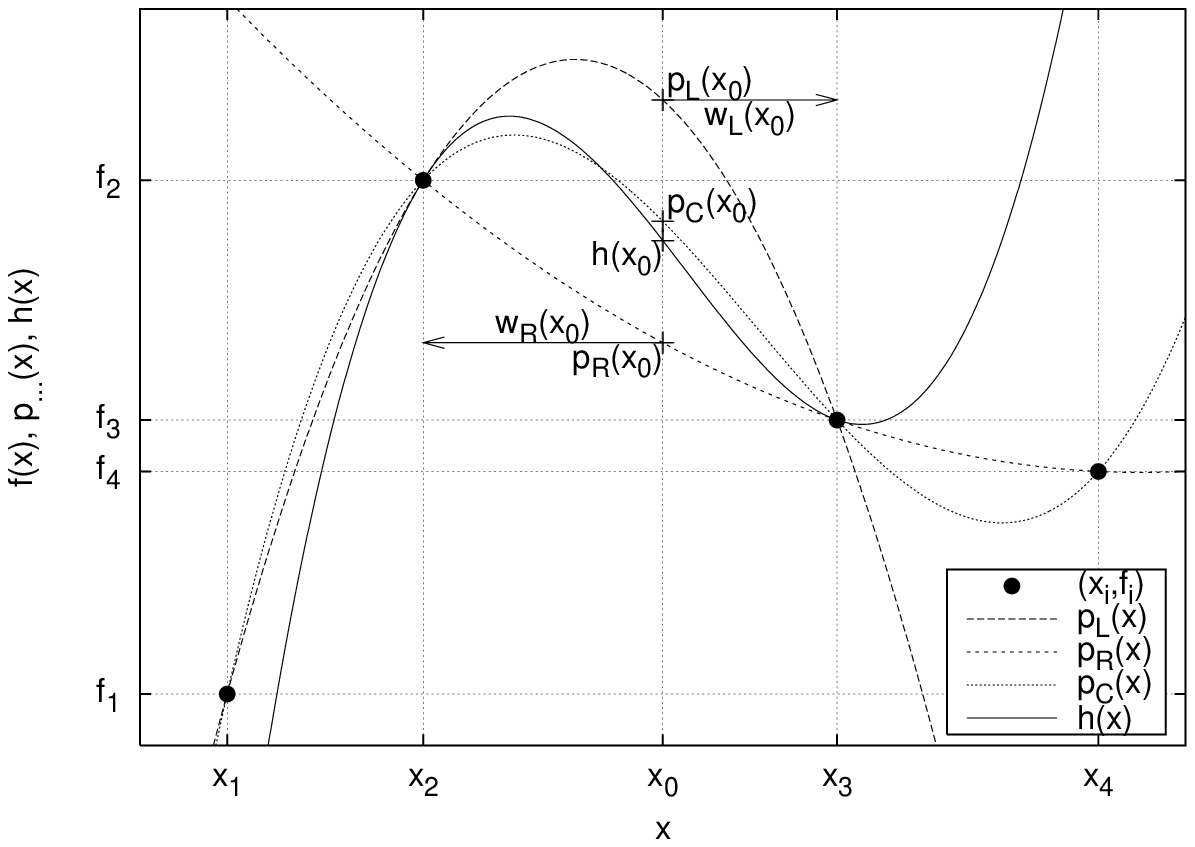}
{Example for distance weighted parabolas as interpolating function compared to a
 cubic polynomial. $h(x)$ is only used for $x = x_0 \in [x_2,x_3]$. Within this
 interval it is quite close to $p_{\rm C}(x)$ while outside the difference
 between the two grows rapidly.}
{figIP}{t}{clip,angle=0,width=115mm}

It can happen that only three neighbors are available (2:1 left:right or 1:2 or
3:0 or 0:3). This is always the case for $x_{(4)} = v =\vmic$, because there
are only four different values in total (0, 1, 2, and 4). In this case, a
simple quadratic (2$^{\sf nd}$ order polynomial) interpolation is done and the
weighting factor is given by:
\be
  w_{(s)[3]} = \frac12 \cdot \frac4{d_{(s)[3]}} = \frac2{d_{(s)[3]}} \le \frac12
\ee
with a factor $1/2$ to account for the lower accuracy of this interpolation due
to one fewer neighbor and $4$ as minimal (optimal) distance sum.

\begin{table}[tb]
 \begin{tabular}{|l|*{6}{r|}}
  \cline{2-7}
     \multicolumn{0}{c}{}\vline
   & \multicolumn{3}{c}{Number of models}\vline
   & \multicolumn{3}{c}{Percentage} \vline \\
  \hline
  \hfill Interpolation     & before &  after &$\Delta$ \ 
                                                  & before& after &$\Delta$ \ \\
  \hline\hline
  Primary\hspace{4.2mm}(fully)
                        & 67,674 & 67,804 & +130  & 73.94 & 74.09 & +0.14\\
  \hline
  Secondary (part.)     & 22,736 & 23,133 & +397  & 24.84 & 25.28 & +0.43\\
  \hline
  Missing / Interpol.     &  1,110 &    583 & $-$527&  1.21 &  0.64 & $-$0.58\\
  \hline\hline
  \hfill Sum & 91,520 & 91,520 & $\pm$ 0 & 100.00 & 100.00 & $\pm$0.00\\
  \hline
 \end{tabular}
 \caption{Numbers and percentages of models fulfilling the primary and
          secondary \smgt\ convergence criteria before and after the
          interpolation. Due to rounding the percentages may not sum up to
          100.00.}
 \label{tabNumCrit}
\end{table}

In case of only two neighbors (2:0, 1:1, or 0:2) linear interpolation is
applied and
\be
  w_{(s)[2]} = \left( \frac12 \right) ^2 \cdot \frac2{d_{(s)[2]}}
             = \frac1{2d_{(s)[2]}} \le \frac14
\ee

With only one neighbor (1:0, 0:1) we use identical interpolation (as an initial
guess) and
\be
  w_{(s)[1]} = \left( \frac12 \right) ^3 \cdot \frac1{d_{(s)[1]}}
             = \frac1{8d_{(s)[1]}} \le \frac18
\ee

If there is no neighbor at all (0:0) we set
\be
  w_{(s)[0]} = 0
\ee

The general expressions for $n=1,\dots,4$ are:
\bea
  \label{eqGeneral}
  w_{(s)[n]} &=& \left(\frac12\right)^{4-n} \cdot 
                  \frac{(-n^3+9n^2-14n+12)/6}{d_{(s)[n]}} \\
  d_{(s)[n]} &=& \sum_{i=1}^n|m_{x_{(s)0}}-m_{x_{(s)i}}| \nn
\eea
  
Once the 1D interpolations in all four dimensions $x_{(s)}$ ($t$, $g$, $z$,
$v$) have been completed, the final result of grid gap closing interpolation is
given by
\bea
  \label{eqFinIp}
  \vec h(\vec x_0) &=& \frac1W \cdot \sum_{s=1}^4 w_{(s)} \cdot
                                                  \vec h_{(s)}(\vec x_0) \\
       q(\vec x_0) &=& \frac W4 \nn\\
       W           &=& \sum_{s=1}^4 w_{(s)} \nn
\eea
Thus, $w_{(s)}$ not only takes into account the distance of neighbors in one
dimension but also the weighting among the 4 dimensions. Strictly speaking it
should be written as $w_{(s)[n_{(s)}(\vec x_0)]}$.

$q(\vec x_0)$ is the {\itshape interpolation/neighbor quality}. The maximum
possible value of it is $1.0$ (100\%) provided that in all four dimensions
there is an optimal 2:2 neighbor situation. In our grid $q(\vec x_0)$ is never
$0.0$,\footnote{That would happen in case of no neighbors at all in all four
dimensions. The maximum value is $87.5\%$ and not $100\%$ because of the
maximal three neighbors in $v$.} hence $\vec h(\vec x_0)$ is always well
defined.\\

\figureDSSN{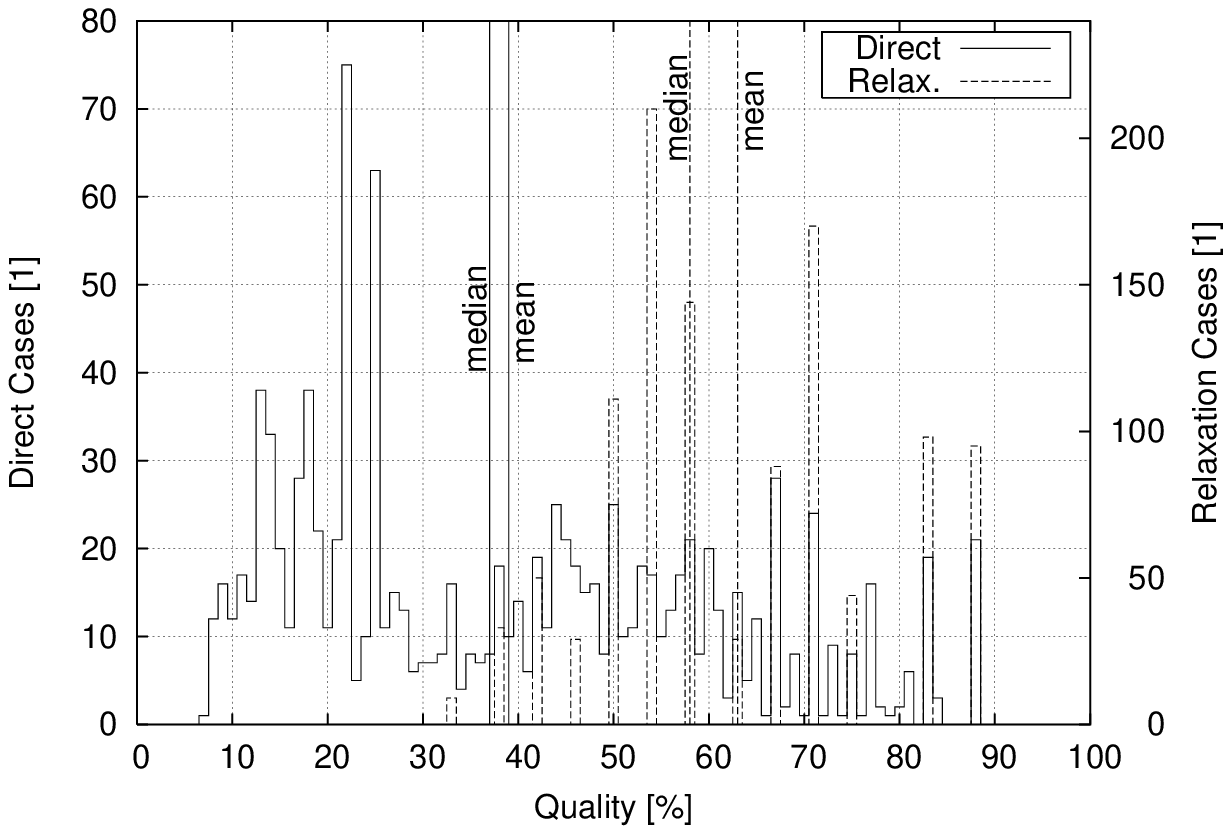}
{Histogram of neighbor/interpolation qualities in the grid with 1,110
 missing/interpolated models. For relaxation interpolation there are only 13
 possible values.}
{figQual}{t}{clip,angle=0,width=115mm}

This procedure (Eq.~\ref{eqMisMod} - \ref{eqFinIp}) is applied to all missing
models in each of the four convection  model subgrids. There are two ways to do
so. We call the first one {\itshape direct interpolation}: make one run for
every missing model as described above and take only converged models as
neighbors. The second one we call {\itshape relaxation interpolation}: make one
run after the other, and take also interpolated models as
neighbors,\footnote{Alternatively, we could save the interpolated files from
the current run elsewhere and replace all the old interpolated models from the
last run with the new ones from the current run only after the current run has
been completed. But as we are not interested in the intermediate results and to
save time a recently interpolated file replaces the old one immediately.} as
long as the (values in the) model files change.\footnote{The comparison is done
character by character (csh-script) which corresponds to a numerical
significance of at most 8 digits, depending on the variable to be interpolated.}

Fig.~\ref{figQual} shows the distribution of the qualities in the grid before
the interpolation. In the case of relaxation interpolation there are no gaps in
the grid anymore, because also the interpolated files are valid neighbors,
hence there are only 13 possible values of neighbor quality (Table
\ref{tab13}).

\begin{table}[t]
 \begin{tabular}{|*{8}{c|}}
  \hline
  Neighbors & \multicolumn{4}{c}{Dimension $s:x_{(s)}$}\vline
                                     & \# & Distance sum & Weight\\
  \hline
  left:right & 1:$t$ & 2:$g$ & 3:$z$ & 4:$v$ & $n$ & $d_{(s)[n]}$
                                                   & $w_{(s)[n]}$\\
  \hline\hline
  3:0, 0:3   & + & + & + & + & 3 & 3+2+1=6
                                  & $\nicefrac12\cdot\nicefrac46=\nicefrac13$ \\
  \hline
  2:1, 1:2   & + & + & + & + & 3 & 2+1+1=4
                                  & $\nicefrac12\cdot\nicefrac44=\nicefrac12$ \\
  \hline
  2:2        & + & + & + &$-$& 4 & 2+1+1+2=6 & $\nicefrac66=1$ \\
  \hline\hline
  ${\displaystyle \sum_{s=1}^4 w_{(s)[n]}}$ &
    \multicolumn{7}{c}
      {$\begin{array}{c}
        \{\nicefrac13,\nicefrac12,1\}_t + \{\nicefrac13,\nicefrac12,1\}_g + 
        \{\nicefrac13,\nicefrac12,1\}_z + \{\nicefrac13,\nicefrac12\}_v =  \\
        =\{1\frac13,1\frac12,1\frac23,1\frac56,2,2\frac16,
           2\frac13,2\frac12,2\frac23,2\frac56,3,3\frac13,3\frac12\}
       \end{array}$}\vline \\
  \hline
  \rule[-1.5mm]{0mm}{4.8mm}$q(\vec x_0)$ [\%]&
    \multicolumn{7}{c}
   {\hspace{-1mm}$\{33\frac13,37\frac12,41\frac23,45\frac56,50,54\frac16,
                    58\frac13,62\frac12,66\frac23,70\frac56,75,83\frac13,
                    87\frac12\}$\makebox[-1mm]{}}\vline \\
  \hline
 \end{tabular}
 \caption{Possible neighbor situations and values of distance sum $d_{(s)[n]}$
          and weighting factor $w_{(s)[n]}$ for the four parameters $x_{(s)}$
          and resulting 13 possible values of the neighbor/interpolation quality
          $q(\vec x_0)$ for relaxation interpolation.}
 \label{tab13}
\end{table}


We save the directly interpolated files and the relaxation interpolated ones in
different directories and apply \smgt\ to them. Different numbers of models
converge (fully and partially) in the two directories. As there are more
converged files in the relaxation directory, which we therefore rate as the
`better' one, we replace the interpolated files in our grid with one of the
appropriate newly converged ones in following order:
\begin{enumerate}
  \item Fully     converged from relaxation interpolated model
  \item Fully     converged from directly   interpolated model
  \item Partially converged from relaxation interpolated model
  \item Partially converged from directly   interpolated model
\end{enumerate}
In this way we can replace 427 (115, 12, 267, and 33 from group 1, 2, 3, and 4,
respectively) interpolated files by converged models.

Fig.~\ref{figRelProg} shows the relaxation progress of the 1,100 missing models
in our grid. 379 runs are needed to converge the procedure.

\figureDSSN{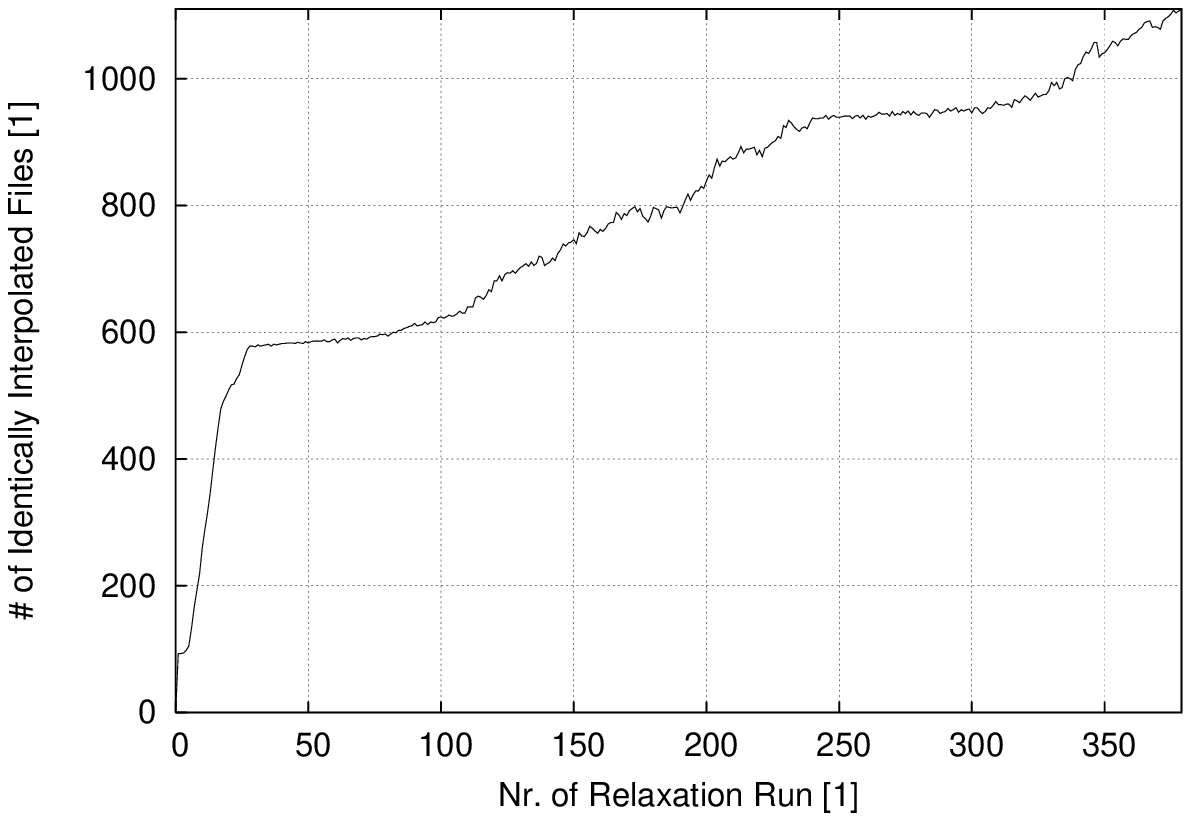}
{Relaxation progress in loop 1: Number of identically interpolated files in
 consecutive relaxation runs versus number of relaxation run (added up over the
 four convection model subgrids). There are actually stages with decreasing
 number. 379 runs are need to converge the procedure for the 1110
 missing/interpolated files which makes an average slope of about 3 files per
 run.}
{figRelProg}{t}{clip,angle=0,width=115mm}

Now with this partially new grid\footnote{Converged files contain different
values compared to the interpolated ones taken as their \smgt\ starting
models.} we can repeat the whole procedure: interpolating directly and
relaxationally, replacing interpolated files with converged models. After 6 such
loops no further progress can be achieved in our grid. A total of 527
(=427+62+20+10+6+2+0 in the 7 loops) out of the originally 1,110 missing models
converged, 583 are left interpolated. (They are called ``missing models after
the interpolation" in Figs.~\ref{figMissTG} - \ref{figMissGZ2}.)


The relaxation progress of the other loops (2 - 7) looks similar to that of
loop 1 (Fig.~\ref{figRelProg}).

328 out of the 527 models converged both starting from a relaxationally
interpolated model and a directly interpolated one, 137 converged only in the
relaxation case and 62 only in the direct one, so without relaxation
interpolation only 390 would have converged (720 left missing/interpolated) and
without direct interpolation 465 (645).

To sum up, the whole grid gap interpolation consists of seven loops, each loop
of one direct interpolation and one relaxation interpolation, each relaxation
interpolation of some hundred runs,\footnote{379, 365, 340, 331, 280, 216, and
166 in the loops 1 to 7, respectively.
} each direct interpolation and each
run consists of four 1D interpolations for every missing model. This gives a
total of 5,893,792 1D interpolations (of vectors/arrays with 6$\cdot$288 = 1728
or 6$\cdot$72 = 432 components). After each loop \smgt\ is applied to the
interpolated files, altogether 4,774 times.


\subsubsection*{Interpolation between grid nodes ({\tt Ip2})}

The {\tt Ip2} is real 4D, it expects a complete $4^4$ (256)
neighborhood\footnote{The base 4 is due to $i=1,\dots,4$ (4 neighbors), the
exponent 4 is due to $s=1,\dots,4$ (4D).} of grid nodes (existing model files,
converged or interpolated)
\be
  \vec f_{ijkl} = \vec f(\vec x_{ijkl})
                = \vec f(x_{(1)i},x_{(2)j},x_{(3)k},x_{(4)l})
                = \vec f(t_i,g_j,z_k,v_l)
\ee
where $i,j,k,l=1,\dots,4$ around the interpolation point $\vec x_0 =
(t_0,g_0,z_0,v_0)$ which now is not a grid node but still we assume $x_{(s)1} <
x_{(s)2} \le x_{(s)0} \le x_{(s)3} < x_{(s)4}$ for $s=1,\dots,4$, i.e.~2
left and 2 right `neighbors', in each of the 4 `directions' (dimensions,
parameters) $x_{(s)}$. 

Using again Lagrange's formalism for polynomial interpolation
(cf.\ Eq.~\ref{eqDelta}, p.~\pageref{eqDelta})
\bea
  \Delta_{0j(s)}    &=& x_{(s)}  - x_{(s)j} \\
  \Delta_{ij(s)}    &=& x_{(s)i} - x_{(s)j} \nn\\
  L_{ijk}(x_{(s)}) &=& \ffDelt{0j(s)}{ij(s)}{0k(s)}{ik(s)}
\eea
16 ($2^4$) left-right combined parabolas can be defined, e.g.\ (cf.\ 
Eq.~\ref{eqLP}, p.~\pageref{eqLP}):
\bea
  p_{\rm LRRL}(\vec x)
  &=& \vec f_{1221}\cdot
              L_{123}(t)\cdot L_{234}(g)\cdot L_{234}(z)\cdot L_{123}(v) + \\
  &+& \vec f_{1222}\cdot
              L_{123}(t)\cdot L_{234}(g)\cdot L_{234}(z)\cdot L_{231}(v) + \nn\\
  &+& \vec f_{1223}\cdot
              L_{123}(t)\cdot L_{234}(g)\cdot L_{234}(z)\cdot L_{312}(v) + \nn\\
  \rule{0mm}{4.5mm}
  &+& \vec f_{1231}\cdot
              L_{123}(t)\cdot L_{234}(g)\cdot L_{342}(z)\cdot L_{123}(v) + \nn\\
  &+& \vec f_{1232}\cdot
              L_{123}(t)\cdot L_{234}(g)\cdot L_{342}(z)\cdot L_{231}(v) + \nn\\
  &+& \vec f_{1233}\cdot
              L_{123}(t)\cdot L_{234}(g)\cdot L_{342}(z)\cdot L_{312}(v) + \nn\\
  \rule{0mm}{4.5mm}
  &+& \vec f_{1241}\cdot
              L_{123}(t)\cdot L_{234}(g)\cdot L_{423}(z)\cdot L_{123}(v) + \nn\\
  &+& \vec f_{1242}\cdot
              L_{123}(t)\cdot L_{234}(g)\cdot L_{423}(z)\cdot L_{231}(v) + \nn\\
  &+& \vec f_{1243}\cdot
              L_{123}(t)\cdot L_{234}(g)\cdot L_{423}(z)\cdot L_{312}(v) + \nn\\
   && \raisebox{0mm}[0mm][0mm]\vdots \nn\\
  &+& \vec f_{3441}\cdot
              L_{312}(t)\cdot L_{423}(g)\cdot L_{423}(z)\cdot L_{123}(v) + \nn\\
  &+& \vec f_{3442}\cdot
              L_{312}(t)\cdot L_{423}(g)\cdot L_{423}(z)\cdot L_{231}(v) + \nn\\
  &+& \vec f_{3443}\cdot
              L_{312}(t)\cdot L_{423}(g)\cdot L_{423}(z)\cdot L_{312}(v) = \nn\\
   && \nn\\
  &=& L_{123}(t) \times \left\{
      L_{234}(g) \times \left[
      L_{234}(z) \times \rule{0mm}{3.3mm} \right. \right. \\ 
   && \times \left(
      L_{123}(v)\cdot \vec f_{1221} +
      L_{231}(v)\cdot \vec f_{1222} +
      L_{312}(v)\cdot \vec f_{1223} \right) + \nn\\
   && \hspace{30.4mm} { } +
      L_{342}(z) \times \nn\\ 
   && \times \left(
      L_{123}(v)\cdot \vec f_{1231} +
      L_{231}(v)\cdot \vec f_{1232} +
      L_{312}(v)\cdot \vec f_{1233} \right) + \nn\\
   && \hspace{30.4mm} { } +
      L_{423}(z) \times \nn\\ 
   && \times \left. \left(
      L_{123}(v)\cdot \vec f_{1241} +
      L_{231}(v)\cdot \vec f_{1242} +
      L_{312}(v)\cdot \vec f_{1243} \right) \right] + \nn\\
   && \hspace{13.15mm} { } + 
      L_{342}(g) \times \left[
      L_{234}(z) \times \rule{0mm}{3.3mm} \right. \nn\\ 
   && \times \left(
      L_{123}(v)\cdot \vec f_{1321} +
      L_{231}(v)\cdot \vec f_{1322} +
      L_{312}(v)\cdot \vec f_{1323} \right) + \nn\\
   && \hspace{30.4mm} { } +
      L_{342}(z) \times \nn\\ 
   && \times \left(
      L_{123}(v)\cdot \vec f_{1331} +
      L_{231}(v)\cdot \vec f_{1332} +
      L_{312}(v)\cdot \vec f_{1333} \right) + \nn\\
   && \hspace{30.4mm} { } +
      L_{423}(z) \times \nn\\ 
   && \times \left. \left(
      L_{123}(v)\cdot \vec f_{1341} +
      L_{231}(v)\cdot \vec f_{1342} +
      L_{312}(v)\cdot \vec f_{1343} \right) \right] + \nn\\
   && \vdots \nn\\
   && \times \left. \left. \left(
      L_{123}(v)\cdot \vec f_{1441} +
      L_{231}(v)\cdot \vec f_{1442} +
      L_{312}(v)\cdot \vec f_{1443} \right) \right] \right\} + \nn\\
   && \vdots \nn\\
  &+& L_{312}(t) \times \left\{
      L_{423}(g) \times \left[
      L_{234}(z) \times \rule{0mm}{3.3mm} \right. \right. \nn\\ 
   && \vdots \nn\\
   && \times \left. \left. \left(
      L_{123}(v)\cdot \vec f_{3441} +
      L_{231}(v)\cdot \vec f_{3442} +
      L_{312}(v)\cdot \vec f_{3443} \right) \right] \right\} \nn
\eea
The complete formula contains 81 ($3^4$) $\vec f_{ijkl}$: $i=1,2,3$, $j=2,3,4$,
$k=2,3,4$, $l=1,2,3$. The order of bracketing ($v$ is innermost) is arbitrary.
The distance (4D hypervolume) weight for this example is
\bea
  w_{\rm LRRL}(\vec x) &=&
      \Delta_{30(1)} \cdot\hspace{1pt}
      \Delta_{02(2)} \cdot\hspace{1pt}
      \Delta_{02(3)} \cdot\hspace{1pt}
      \Delta_{30(4)}      \hspace{1pt} = \\
  &=& w_{\rm L}(t) \,\cdot\,
      w_{\rm R}(g) \,\cdot\,
      w_{\rm R}(z) \,\cdot\,
      w_{\rm L}(v) = \nn\\
  &=& (t_3-t)(g-g_2)(z-z_2)(v_3-v) \nn
\eea
The general expressions are given by
\bea
  p_{\hat i\hat j\hat k\hat l}(\vec x) &=& 
    \sum_{\ i=1+\hat i}^{3+\hat i}
    \sum_{\ j=1+\hat j}^{3+\hat j}
    \sum_{\ k=1+\hat k}^{3+\hat k}
    \sum_{\ l=1+\hat l}^{3+\hat l}
    \vec f_{ijkl} \times \\
  &\times& \nn
    \ub{\mathop{\prod_{i'=1+\hat i}^{3+\hat i}}_{i' \ne i\ \ \ }
        \fDelt{0i'(1)}{ii'(1)}}_{L_{ii'_1i'_2}(t)}
    \ub{\mathop{\prod_{j'=1+\hat j}^{3+\hat j}}_{j' \ne j\ \ \ }
        \fDelt{0j'(2)}{jj'(2)}}_{L_{jj'_1j'_2}(g)}
    \ub{\mathop{\prod_{k'=1+\hat k}^{3+\hat k}}_{k' \ne k\ \ \ }
        \fDelt{0k'(3)}{kk'(3)}}_{L_{kk'_1k'_2}(z)}
    \ub{\mathop{\prod_{l'=1+\hat l}^{3+\hat l}}_{l' \ne l\ \ \ }
        \fDelt{0l'(4)}{ll'(4)}}_{L_{ll'_1l'_2}(v)} = \\
  &=& \nn
    \sum_{i=1+\hat i}^{3+\hat i} L_{ii'_1i'_2}(t) \times \left\{
    \sum_{j=1+\hat j}^{3+\hat j} L_{jj'_1j'_2}(g) \times {} \right. \\
  && \nn \times \left[ \left.
    \sum_{k=1+\hat k}^{3+\hat k} L_{kk'_1k'_2}(z) \times \left(
    \sum_{l=1+\hat l}^{3+\hat l} L_{ll'_1l'_2}(v) \times \vec f_{ijkl}
    \right) \right] \right\} \\  \nn\\
  w_{\hat i\hat j\hat k\hat l}(\vec x) &=& 
    w_{\hat i}(t) \cdot w_{\hat j}(g) \cdot w_{\hat k}(z) \cdot w_{\hat l}(v).
\eea
where $\hat i$ reaches the value ${\rm L} = 0$ in case of a left parabola (in
$x_{(1)}=t$) and ${\rm R} = 1$ in case of a right one, and analogously for
$\hat j$ ($x_{(2)}=g$), $\hat k$ ($x_{(3)}=z$), and $\hat l$ ($x_{(4)}=v$), and
\bea
  w_{\rm L}(x_{(s)}) &=& \Delta_{30(s)} \\
  w_{\rm R}(x_{(s)}) &=& \Delta_{02(s)} \nn\\
  W                  &=& 
    \sum_{\hat i={\rm L,R}}
    \sum_{\hat j={\rm L,R}}
    \sum_{\hat k={\rm L,R}}
    \sum_{\hat l={\rm L,R}}
    w_{\hat i\hat j\hat k\hat l}(\vec x) = \\
  &=& \prod_{s=1}^4 \Delta_{32(s)} = 
      (t_3-t_2)(g_3-g_2)(z_3-z_2)(v_3-v_2) \nn
\eea
For the interpolating function we obtain (cf.\ Eq.~\ref{eqDefH},
p.~\pageref{eqDefH})
\bea
  \vec h(\vec x) &=& \frac1W
    \sum_{\hat i=0}^1
    \sum_{\hat j=0}^1
    \sum_{\hat k=0}^1
    \sum_{\hat l=0}^1
    w_{\hat i\hat j\hat k\hat l}(\vec x) \cdot
    p_{\hat i\hat j\hat k\hat l}(\vec x) = \\
  &=& \frac1W
    \sum_{\hat i=0}^1 w_{\hat i}(t)
    \sum_{\hat j=0}^1 w_{\hat j}(g)
    \sum_{\hat k=0}^1 w_{\hat k}(z)
    \sum_{\hat l=0}^1 w_{\hat l}(v) \cdot
    p_{\hat i\hat j\hat k\hat l}(t,g,z,v) = \nn\\
  &=& \frac1W
    \sum_{\hat i=0}^1 w_{\hat i}(t)
    \sum_{\hat j=0}^1 w_{\hat j}(g)
    \sum_{\hat k=0}^1 w_{\hat k}(z)
    \sum_{\hat l=0}^1 w_{\hat l}(v) \times \nn\\
   && \times
    \sum_{i=1+\hat i}^{3+\hat i} L_{ii'_1i'_2}(t) 
    \sum_{j=1+\hat j}^{3+\hat j} L_{jj'_1j'_2}(g) 
    \sum_{k=1+\hat k}^{3+\hat k} L_{kk'_1k'_2}(z)
    \sum_{l=1+\hat l}^{3+\hat l} L_{ll'_1l'_2}(v) \times \nn\\
   && \times \ \vec f_{ijkl} \ = \nn\\
  &=& \frac1W
    \sum_{\hat i=0}^1            w_{\hat i}(t)
    \sum_{i=1+\hat i}^{3+\hat i} L_{ii'_1i'_2}(t) \cdot
    \sum_{\hat j=0}^1            w_{\hat j}(g)
    \sum_{j=1+\hat j}^{3+\hat j} L_{jj'_1j'_2}(g) \times \nn\\
   && \times
\ub{\sum_{\hat k=0}^1            w_{\hat k}(z)
    \sum_{k=1+\hat k}^{3+\hat k} L_{kk'_1k'_2}(z) \cdot
\ub{\sum_{\hat l=0}^1            w_{\hat l}(v)
    \sum_{l=1+\hat l}^{3+\hat l} L_{ll'_1l'_2}(v)
    \cdot \vec f_{ijkl}}_{\vec H_{ijk}(v)}}_{\vec H_{ij}(z,v)} \nn
\eea
This can be transformed into a sum of (tetracubic) separable expressions again
by the definitions
\bea
  \bar L_1(x) &=&        w_{\rm L}(x) \cdot L_{123}(x) \\
  \bar L_2(x) &=&        w_{\rm L}(x) \cdot L_{231}(x) + \nn
                         w_{\rm R}(x) \cdot L_{234}(x) \\
  \bar L_3(x) &=&        w_{\rm L}(x) \cdot L_{312}(x) + \nn
                         w_{\rm R}(x) \cdot L_{342}(x) \\
  \bar L_4(x) &=& \hspx{$w_{\rm L}(x) \cdot L_{312}(x) + {}$} 
                         w_{\rm R}(x) \cdot L_{423}(x) \nn
\eea
And so
\bea
  \vec H_{ijk}(v) &=&
    \sum_{\hat l=0}^1            w_{\hat l}(v)
    \sum_{l=1+\hat l}^{3+\hat l} L_{ll'_1l'_2}(v)
    \cdot \vec f_{ijkl} = \\
  &=& \nn
  \left[ w_0(v) \sum_{l=1}^3 \;+\; 
         w_1(v) \sum_{l=2}^4 \right] L_{ll'_1l'_2}(v) \cdot \vec f_{ijkl} = \\
  &=& w_{\rm L}(v) \left\{ L_{123}(v) \vec f_{ijk1} +
                           L_{231}(v) \vec f_{ijk2} +
                           L_{312}(v) \vec f_{ijk3} \right\} + \nn\\
  &+& w_{\rm R}(v) \left\{ L_{234}(v) \vec f_{ijk2} +
                           L_{342}(v) \vec f_{ijk3} +
                           L_{423}(v) \vec f_{ijk4} \right\} = \nn\\
  &=& \nn
  \vec f_{ijk1} \ub{        w_{\rm L}(v) L_{123}(v)         }_{\bar L_1(v)} + \\
  &+& \nn
  \vec f_{ijk2} \ub{\left\{ w_{\rm L}(v) L_{231}(v) +
                            w_{\rm R}(v) L_{234}(v) \right\}}_{\bar L_2(v)} + \\
  &+& \nn
  \vec f_{ijk3} \ub{\left\{ w_{\rm L}(v) L_{312}(v) +
                            w_{\rm R}(v) L_{342}(v) \right\}}_{\bar L_3(v)} + \\
  &+& \nn
  \vec f_{ijk4} \ub{        w_{\rm R}(v) L_{423}(v)         }_{\bar L_4(v)} = \\
  &=& \sum_{l=1}^4 \bar L_l(v) \cdot \vec f_{ijkl}
\eea
Analogously\footnote{In {\tt Fortran90} the (one line) command for $\vec
H_{ijk}(v) \rightarrow \vec H_{ij}(z,v)$ is e.g.: {\tt hij(:,:,:) =
Lz(1)*hijk(:,:,:,1) + Lz(2)*hijk(:,:,:,2) + Lz(3)*hijk(:,:,:,3) +
Lz(4)*hijk(:,:,:,4)}, where the first `:' stands for the $\vec{\ }$ of $\vec
H$, and the second and third for $i$ and $j$, respectively; no {\tt DO} loops
have to be written for this. $k=1,\dots,4$ is written explicitly.}
\bea
  \vec H_{ij}(z,v) &=&
    \sum_{\hat k=0}^1            w_{\hat k}   (z)
    \sum_{k=1+\hat k}^{3+\hat k} L_{kk'_1k'_2}(z) \cdot
    \vec H_{ijk}(v) = \\
    &=& \sum_{k=1}^4 \bar L_k(z) \cdot \vec H_{ijk}(v) \nn\\
  \vec H_{i}(g,z,v) &=&
    \sum_{\hat j=0}^1            w_{\hat j}   (g)
    \sum_{j=1+\hat j}^{3+\hat j} L_{jj'_1j'_2}(g) \cdot
    \vec H_{ij}(z,v) = \\
    &=& \sum_{j=1}^4 \bar L_j(g) \cdot \vec H_{ij}(z,v) \nn\\
  \vec H(t,g,z,v) &=&
    \sum_{\hat i=0}^1            w_{\hat i}   (t)
    \sum_{i=1+\hat i}^{3+\hat i} L_{ii'_1i'_2}(t) \cdot
    \vec H_{i}(g,z,v) = \\
    = \vec H(\vec x) &=& \sum_{i=1}^4 \bar L_i(t) \cdot \vec H_{i}(g,z,v) \nn\\
  \vec h(\vec x) &=& \frac1W \vec H(\vec x)
\eea
Thus, we finally can write (cf.\ Eq.~\ref{eqIp1D}, p.~\pageref{eqIp1D})
\bea
  \vec h(\vec x)
  &=&
  \frac1W
  \sum_{i=1}^4 \sum_{j=1}^4 \sum_{k=1}^4 \sum_{l=1}^4 \vec f_{ijkl}
  \bar L_i(t)  \bar L_j(g)  \bar L_k(z)  \bar L_l(v) = \\
  &=&
  \frac1W
  \ub { \sum_{i=1}^4 \bar L_i(t)
  \ub { \sum_{j=1}^4 \bar L_j(g)
  \ub { \sum_{k=1}^4 \bar L_k(z)
  \ub { \sum_{l=1}^4 \bar L_l(v) \vec f_{ijkl} }_{\vec H_{ijk}  (v)}
                                               }_{\vec H_{ij} (z,v)}
                                               }_{\vec H_{i}(g,z,v)}
                                               }_{\vec H  (t,g,z,v)} \nn
\eea
The separability is of great advantage for the differentiation, e.g.\ ($s=2$):
\bea
  \frac{\partial\vec h(\vec x)}{\partial g}
  &=&
  \frac1W
  \sum_{i=1}^4 \sum_{j=1}^4                        \sum_{k=1}^4 \sum_{l=1}^4
  \vec f_{ijkl}
  \bar L_i(t)  \frac{{\rm d}\bar L_j(g)}{{\rm d}g} \bar L_k(z)  \bar L_l(v) = \\
  &=&
  \frac1W
  \ub { \sum_{i=1}^4 \bar L_i(t)
  \ub { \sum_{j=1}^4 \bar L'_j(g)
  \ub { \sum_{k=1}^4 \bar L_k(z)
  \ub { \sum_{l=1}^4 \bar L_l(v) \vec f_{ijkl} }_{\vec H_{ijk}          (v)}
                                               }_{\vec H_{ij}         (z,v)}
                                               }_{\vec H_{i}^{(g)}  (g,z,v)}
                                               }_{\vec H    ^{(g)}(t,g,z,v)} \nn
\eea
with
\bea
  w'_{\rm L}(x) &=& -1 \\
  w'_{\rm R}(x) &=& +1 \nn\\
  L'_{ijk}(x)   &=& \frac{\Delta_{0j} +   \Delta_{0k}}
                         {\Delta_{ij}\cdot\Delta_{ik}} \\
  \bar L'_1(x) &=&     w_{\rm L}(x)\cdot L'_{123}(x) 
           \hspx{${} + w_{\rm R}(x)\cdot L'_{234}(x)$} - L_{123}(x)\\
  \bar L'_2(x) &=&     w_{\rm L}(x)\cdot L'_{231}(x) + \nn
                       w_{\rm R}(x)\cdot L'_{234}(x) - L_{231}(x) + L_{234}(x)\\
  \bar L'_3(x) &=&     w_{\rm L}(x)\cdot L'_{312}(x) + \nn
                       w_{\rm R}(x)\cdot L'_{342}(x) - L_{312}(x) + L_{342}(x)\\
  \bar L'_4(x) &=& \hspx{$w_{\rm L}(x)\cdot L'_{312}(x) + {}$} \nn 
                          w_{\rm R}(x)\cdot L'_{423}(x) 
                                         \hspx{${} - L_{312}(x)$} + L_{423}(x)
\eea
Derivatives of order higher than one (which are not continuous at the grid
nodes!) and mixed derivatives can also be computed easily in this way.

If $x_{(s)0} = x \in [x_1,x_2]$ or $x \in [x_{N-1},x_N]$ (no two neighbors on
each side) then simple quadratic interpolation is applied:
$p_Q$($x$; $x_1$,$x_2$,$x_3$) or $p_Q$($x$; $x_{N-2}$,$x_{N-1}$,$x_N$).

For grids with fewer or more (numerical) parameters/dimensions $x_{(s)}$ the
interpolation works analogously.\footnote{The number of neighbors will not
change: $i=1,\dots,4$, but the number of dimensions/'directions':
$s=1,\dots,D$.} 


\section*{Discussion and outlook}


\subsection*{Improvement of convergence}

The rate of success of the grid gap interpolation procedure described in the
previous section depends on the location in the parameter space. By success we
mean that newly converged models are obtained when using the interpolated ones
as starting models. From Figs.\ \ref{figMissT} to \ref{figMissC} one can see
that the procedure removed most of the isolated gaps for parameters where most
of the surrounding models are converged (i.e.\ with high neighbor quality). It
was also successful for a part of the transition region in the HR diagram where
convection switches from thin superadiabatic zones as in A stars to deep
quasi-adiabatic zones as in solar type stars. This region extends along a line
from around 5\,500~K at $\logg=2$ to around 7\,500~K for $\logg=5$ and shifts to
slightly higher temperatures for increasing metallicity. For the part of this
region with high \logg, the interpolation presumably produces models which are
close enough to the `correct' structure so that the subsequent model
calculation is safe from temporarily converging to a wrong solution. The gap
remaining at the low \logg\ end of this region (for all metallicities
considered here) indicates that the interpolation is not accurate enough to
overcome this `transition problem' there.

Interpolation is also ineffective in the corner of our parameter space
characterized by low metallicity, low temperature and high \logg, where the
temperature correction algorithm applied in \atlas cannot cope with adiabatic
convection in the atmosphere. The third region where models remain missing
after the interpolation concerns models with 288 layers and $\mh=+1.0$. The
region is confined to temperatures between 8\,400 and 8\,800~K and \logg\ larger
than 3.6. For those models the flux derivative errors high in the atmosphere
($\logtau \lesssim -5$) do not decrease below 10\% and continuously increase to
about 40\% in the outermost layers. The flux errors and temperature corrections
are very small (lower than the primary convergence criteria). The same
phenomenon occurs in the models with 72 layers, but is confined to even higher
layers ($\logtau\ \lesssim -6$) and therefore they have been classified as
converged according to the secondary criteria. We cannot give a definitive
explanation for this problem at the moment.

For a part of the models in the grids the gas pressure decreases to values
below $10^{-2}$~dyn~cm$^{-2}$ for the outermost layers. These values are
outside of the range of the ODF tables by Kurucz (1993a), and \atlas uses the
constant value corresponding to the limiting pressure instead. For models with
$\mh=-2.0$ this is the case for a nearly triangular region with $\teff \ge
6\,500$~K and $\logg \le 4.2$, where at these limits only the uppermost layer
is affected, while the extension of this zone increases gradually towards the
inner layers when moving towards the `corner' of our parameter space. In the
most extreme case (\teff, \logg) = (10\,000, 2.2) it extends to $\logtau
\lesssim -4.5$ and the pressure at the outermost layer is $\approx
10^{-4.5}$~dyn~cm$^{-2}$. The affected region is smaller for higher
metallicities. This inaccurate description of the opacity at high layers
nevertheless does not seem to present a problem for the convergence of these
models, as the temperature corrections, flux and flux derivative errors do not
show a correlated behavior.\footnote{See the figures at the \nemo\ website
http://ams.astro.univie.ac.at/nemo/dvd/??????/vmicro\_?/, where '??????' is one
of 'cgm072', 'cgm288', 'mlt072', or 'cm\_288', and '?' is one of 0,1,2,4; files
`info\_p.ps' and `info\_m.ps'.} The only exception might be the $\mh=-2$
subgrids where the hottest models show increasingly larger temperature
corrections for low \logg\ (but small flux and flux derivative errors).


\subsection*{Interpolation routines}

At present the interpolation routines work only with the \nemo\ grid and
\atlas\ atmospheres with 72 or 288 layers. Future versions are intended to deal
with a variety of different grid structures and resolutions in a flexible
manner.


\acknowledgments{This work is supported by the Austrian Fonds zur F\"orderung
der wissenschaftlichen Forschung (FWF) within the project Stellar Atmospheres
and Pulsating Stars (P14984), by the Bundesmininsterium f\"ur Bildung,
Wissenschaft und Kultur (BMBWK) and the Bundesmininsterium f\"ur Verkehr,
Innovation und Technologie (BMVIT) via the Austrian Space Agency (ASA).\\
Many thanks to Barry Smalley for his numerous advices and corrections!}

\References{

Arp,~H. 1961, A.J. 133, 874\\

A$\check{\sf z}$usienis,~A., Strai$\check{\sf z}$ys,~V. 1969, Soviet Astron. 13,
316\\

Barban,~C., Goupil,~M.~J., van't~Veer-Menneret,~C., Garrido,~R., Kupka,~F.,
\ri Heiter,~U. 2003, A\&A 405, 1095\\

Buser,~R. 1978, A\&A 62, 411\\

Buser,~R., Kurucz,~R.~L. 1978, A\&A 70, 555\\

Cousins,~A.~W.~J. 1976, Mem. R. Astron. Soc. 81, 25\\

Cramer,~N., Maeder,~A. 1979, A\&A 78, 395\\

Crawford,~D.~L., Barnes,~J.~V. 1970, A.J. 75, 978\\

D'Antona,~F., Montalb\'an,~J., Kupka,~F., Heiter,~U. 2002, ApJ 564, L93\\

D'Antona,~F., Montalb\'an,~J. 2003, A\&A 412, 213\\

Garrido,~R., Garcia-Lobo,~E., Rodriguez,~E. 1990, A\&A 234, 262\\

Garrido,~R., Claret,~A., Moya,~A., Kupka,~F., Heiter,~U., Barban,~C.,
Goupil,~M.~J.,
\ri van't~Veer-Menneret,~C. 2001, in Proc. 1st Eddington workshop
\ri {\em Stellar Structure and Habitable Planet Finding}, ed. F.~Favata,
I.W.~Roxburgh
\ri  and D.~Galad\`i (ESA SP-485, Noordwijk: ESA Publications
Division), 103\\

Garrido,~R., Moya,~A., Goupil,~M.~J., Barban,~C., van't~Veer-Menneret,~C.,
\ri Kupka,~F., Heiter,~U. 2002, CoAst 141, 48\\


Hauck,~B., Mermilliod,~M. 1980, A\&AS 40, 1\\

Heiter,~U., van't~Veer-Menneret,~C., Barban,~C., Weiss,~W.~W., Goupil,~M.~J.,
\ri Schmidt~W., Katz,~D., Garrido,~R. 2002, A\&A 392, 619\\


K\"unzli,~M., North,~P., Kurucz,~R.~L., Nicolet,~B. 1997, A\&AS 122, 51\\

Johnson,~H.~L., Mitchell,~R.~I. 1975, Rev. Mex. Astron. Astrofis. 1, 299\\

Kupka~F., Paunzen~E., Maitzen~H.~M. 2003, MNRAS 341, 849-854\\

Kurucz,~R.~L. 1993a, {\em Opacities for Stellar Atmospheres}, Kurucz CD-ROMs
No.~2--6,
\ri Cambridge, Mass.: Smithsonian Astrophysical Observatory\\

Kurucz,~R.~L. 1993b, {\em ATLAS9 Stellar Atmosphere Programs and 2 km/s grid},
\ri Kurucz CD-ROM No.~13, Cambridge, Mass.: Smithsonian
\ri Astrophysical Observatory\\

Kurucz,~R.~L. 1998, http://kurucz.harvard.edu/\\

Lamla,~E. 1982, in Landolt-B{\"o}rnstein New Series, vol. VI 2b, ed.
K.~Schaifers
\ri and H.~H.~Voigt (Berlin: Springer), p.35-90\\

Lub,~J., Pel,~J.~W. 1977, A\&A 54, 137\\

Maitzen,~H.~M., Vogt,~N. 1983 A\&A 123, 48\\

Matthews,~Th.~A., Sandage,~A.~R. 1963, A.J. 138, 30\\



Montalb\'an,~J., D'Antona,~F., Kupka,~F., Heiter,~U. 2004, A\&A 416, 1081\\

Nendwich,~J., Nesvacil,~N., Heiter,~U., Kupka,~F. 2003, in IAU Symp. 210,
\ri {\em Modelling of Stellar Atmospheres}, ed. N.E.~Piskunov, W.W.~Weiss
\ri and D.F.~Gray (San Francisco: ASP), p.~423\\

North,~P. private communication with F.~Kupka, 1997\\

Rufener,~F., Nicolet,~B. 1988, A\&A 206, 357\\

Schuler,~S.~C., King,~J.~R., Hobbs,~L.~M., Pinsonneault,~M.~H. 2004, ApJ 602,
L117\\

Smalley,~B., Gardiner,~R.~B., Kupka,~F., Bessell,~M.~S. 2002, A\&A 395, 601\\

Smalley,~B., Kupka,~F. 2003, in IAU Symp. 210, {\em Modelling of Stellar
Atmospheres},
\ri ed. N.E.~Piskunov, W.W.~Weiss and D.F.~Gray (San Francisco: ASP), p.~425\\

Stassun,~K.~G., Mathieu,~R.~D., Vaz,~L.~P.~R., Stroud,~N., Vrba,~F.~J. 2004,
\ri ApJS 151, 357\\

Str\"omgren,~B. 1966, Ann. Rev. Astron. Astrophys. 4, 433\\


}

\end{document}